\documentclass[numberedappendix,twocolumn,iop]{emulateapj} 

\usepackage{latexsym}   
\usepackage{amsfonts}   
\usepackage{amstext}    
\usepackage{amsmath,amssymb}
\usepackage{mathtools}
\usepackage{braket}
\usepackage{bm}
\usepackage{empheq} 
\usepackage{MnSymbol} 

\usepackage{subfigure}

\slugcomment{}

\begin{document}
\title{Spatio-temporal evolution of Hanle and Zeeman \\ synthetic polarization in a chromospheric spectral line}

 \author{E.S. Carlin\altaffilmark{1}, Bianda M.\altaffilmark{1}}
\altaffiltext{1}{Istituto Ricerche Solari Locarno, 6600, Locarno, Switzerland}  
\email{escarlin@irsol.es}


\begin{abstract}
Due to the quick evolution of the solar chromosphere, its magnetic
field cannot be inferred reliably without accounting for the temporal
variations of its polarized light. This has been broadly overlooked in the
modelling and interpretation of the polarization due to technical
problems (e.g., lack of temporal resolution or of time-dependent MHD
solar models) and/or because many polarization measurements can
apparently be explained without dynamics. Here, we show that the
temporal evolution is critic for explaining the spectral-line
scattering polarization because of its sensitivity to rapidly-varying
physical quantities and the possibility of signal cancellations and
attenuation during extended time integration. For studying the combined effect of  time-varying magnetic
fields and kinematics, we solved the $1.5$D NLTE problem of the second
kind in time-dependent 3D R-MHD solar models and we synthesized the
Hanle and Zeeman polarization in forward scattering for the chromospheric $\lambda4227$ line. We find that the
quiet-sun polarization amplitudes depend on the periodicity and
spectral coherence of the signal enhacements produced by kinematics,
but that substantially larger linear polarization signals should exist
all over the solar disk for short integration times. The spectral
morphology of the polarization is discussed as a combination of
Hanle, Zeeman, dynamic effects and partial redistribution. We give
physical references for observations by degrading and characterizing
our slit time-series in different spatio-temporal resolutions. The
implications of our results for the interpretation of the second solar
spectrum and for the investigation of the solar atmospheric heatings
are discussed.

\end{abstract}


\section{Introduction}
Two 4-m class solar
telescopes (EST: \cite{Collados:2013aa}; and DKIST:
\cite{Rimmele:2013aa}) with exceptional spectropolarimetric
capabilities are being developed at the present moment. 
They are expected to provide a
 sensitivity of $10^{-4}$  
while preserving the spatio-temporal resolution of $\approx
0.^{\prime\prime}1 \times 20$
s required for following the
evolution of the chromospheric spectral line polarization.
 Without it the signals end up being significantly integrated in
 space, time and/or wavelength, either intrinsically, by an instrument without
 enough resolution, or after detection, for increasing the S/N
 delivered by the spectropolarimeter.

Thus, spatiotemporal integration limits the study of the quiet solar chromosphere.
Possible sign cancellations below the resolution element definitely kills the
already faint transversal Zeeman signals produced by the weak chromospheric
magnetism. The only alternative are then scattering signals, which can exist
 even in absence of magnetic field.
Atoms scattering polarized light is the closest thing to an
ideal in-situ detector of plasma properties (with data transponder
incorporated), exhibiting
large responsiveness to chromospheric
magnetic fields by Hanle effect, but also to the
 chromospheric temperature  and velocity gradients via changes in radiation
field anisotropy and atomic polarization
\citep[][]{carlin12,Carlin:2013aa}. Anisotropic
  radiation also adds sensitivity to the horizontal inhomogeneities in the
plasma when the three-dimensional structure of the radiation field is
considered \citep{Stepan:2015aa,Tichy:2015aa}. When by this or other reason the symmetry in the scattering is broken (e.g., by a non-radial
magnetic field in forward scattering), the modulation
of chromospheric atomic polarization produced by shock waves becomes
visible in large and frequent changes in the
shape, sign and amplitude of the
emergent profiles
 \citep[][]{Carlin:2015aa}. Thus,
it has been recently pointed out
that in contrast with previous expectations sign cancellations can also affect the Hanle polarization
signatures for current typical resolutions \citep[][hereafter the 
Letter]{Carlin:2016aa}. 

Note that the current maximum \textit{spatial} resolution seems enough for tracking
spatial variations of the quiet chromospheric magnetism through
scattering Hanle signals (which is yet harder with the
transversal Zeeman effect). However, the temporal
scales of the chromosphere are significantly shorter than the
several minutes that most
observations of scattering polarization last. Hence, analyzing
  a time average the information contained in the temporal
  evolution -e.g. continuity and causality of events- is lost, and the
  comparison with calculations becomes misleading. 
Furthermore, as
chromospheric events can be very fast, they are not statistically
well represented in calculations with single MHD snapshots of limited
extension.
For target resolutions around $0.^{\prime\prime}1 \times 20$ s,
our calculations
 support detection of \textit{near
ubiquitous} scattering polarization signals in the quiet 
chromosphere once a sensitivity of $10^{-4}$ ($10^{-3}$ for some spectral lines) 
 is surpassed.
This threshold
is quiet particular. Its crossing is a sort of
discrete leap that should allow the detection of the faint disk-center quiet sun
polarization, thus obtaining an almost fully polarized solar disk in several
spectral lines, something to expect with the coming generation of solar facilities. 
Thus, our simulations try to estimate the polarization that a 4-m
class solar telescope might observe when the dynamic signals are
measured in their proper time scales. On the other hand, we
expect that this work help to find a way of disentangling the effect of
velocities and magnetic fields in the Hanle signals of chromospheric lines. 
Our recent Letter advanced some of the results, pointing out that a
minimal understanding of the temporal evolution of the polarization is required for determining magnetic fields in dynamic layers as well as
for deciphering the second solar spectrum.

In the present paper we continue studying the
Stokes signals of the
$\lambda4227$ line (located at $4226.728$ {\AA} in air). This
paradigmatic spectral line of the second solar spectrum has been widely
studied, both observationally and theoretically, in the last $50$
years. Some examples are \cite{bruckner1963}, \cite{Dumont:1973aa},
\cite{Stenflo:1974aa}, \cite{Dumont:1977aa},
\cite{Faurobert-Scholl:1992aa},
\cite{Bianda:1998aa}, \cite{Holzreuter:2005aa}, \cite{Sampoorna:2009aa},
\cite{Anusha:2010aa}, \cite{michele2011cai} or \cite{Supriya:2014aa}.
We improve upon these
studies by using a time-series of realistic 3D radiation-MHD models as input atmospheres, and by studying
the effect of spatio-temporal integration and dynamics on the Hanle
and Zeeman signals in the whole Stokes vector. 
We call Ca {\sc i} $\mathrm{4227 \,
  {\AA}}$ a \textit{reference line }because the quantification of dynamic
effects in its spectral core gives useful physical insights for unpuzzling other
scattering signals forming at similar heights but with richer 
atomic structure, such as the Na{\sc i} D lines. $\lambda\mathrm{4227}$ seems indeed ideal for this purpose because it is
a chromospheric spectral line with
minimal quantum complexity: normal Zeeman
triplet, no hyperfine structure, no lower level
polarization. Furthermore, its
large forward-scattering polarization signals in all Stokes parameters
permits to explore the lower chromosphere at and around 
disk center, which avoids the more complex interpretation of line-of-sight
superposition effects at the solar limb \citep[see introduction of][]{Carlin:2015ab}. 

After presenting our results (Sec. \ref{sec:results}), we present some discussions
(sec. \ref{sec:unexplained}). The key in Sec. \ref{sec:unexplained}
is that the degeneracy in the solar signals can lead to close fits
with simulations implying, however, wrong physical inferences; and that this is avoided with larger spatio-temporal
resolutions and a precise characterization of the effect of
chromospheric dynamics (i.e., time evolution of macroscopic motions and heatings) in the polarization.
\section{Synthesis of the polarization signals}\label{sec:calcs} 
\begin{figure*}[t!]
\centering$
\begin{array}{cc} 
\includegraphics[scale=0.7]{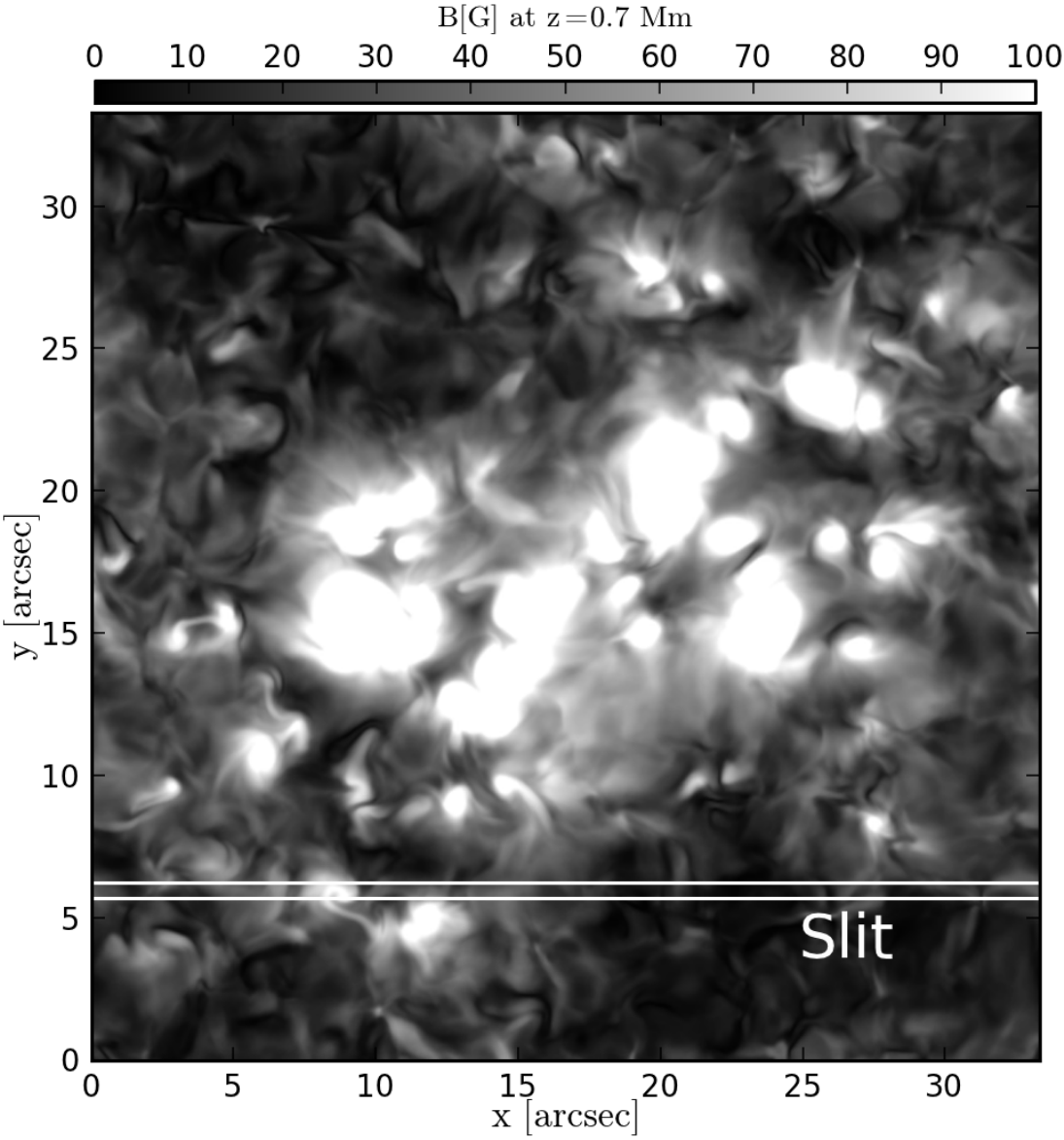}   & 
\includegraphics[scale=0.7]{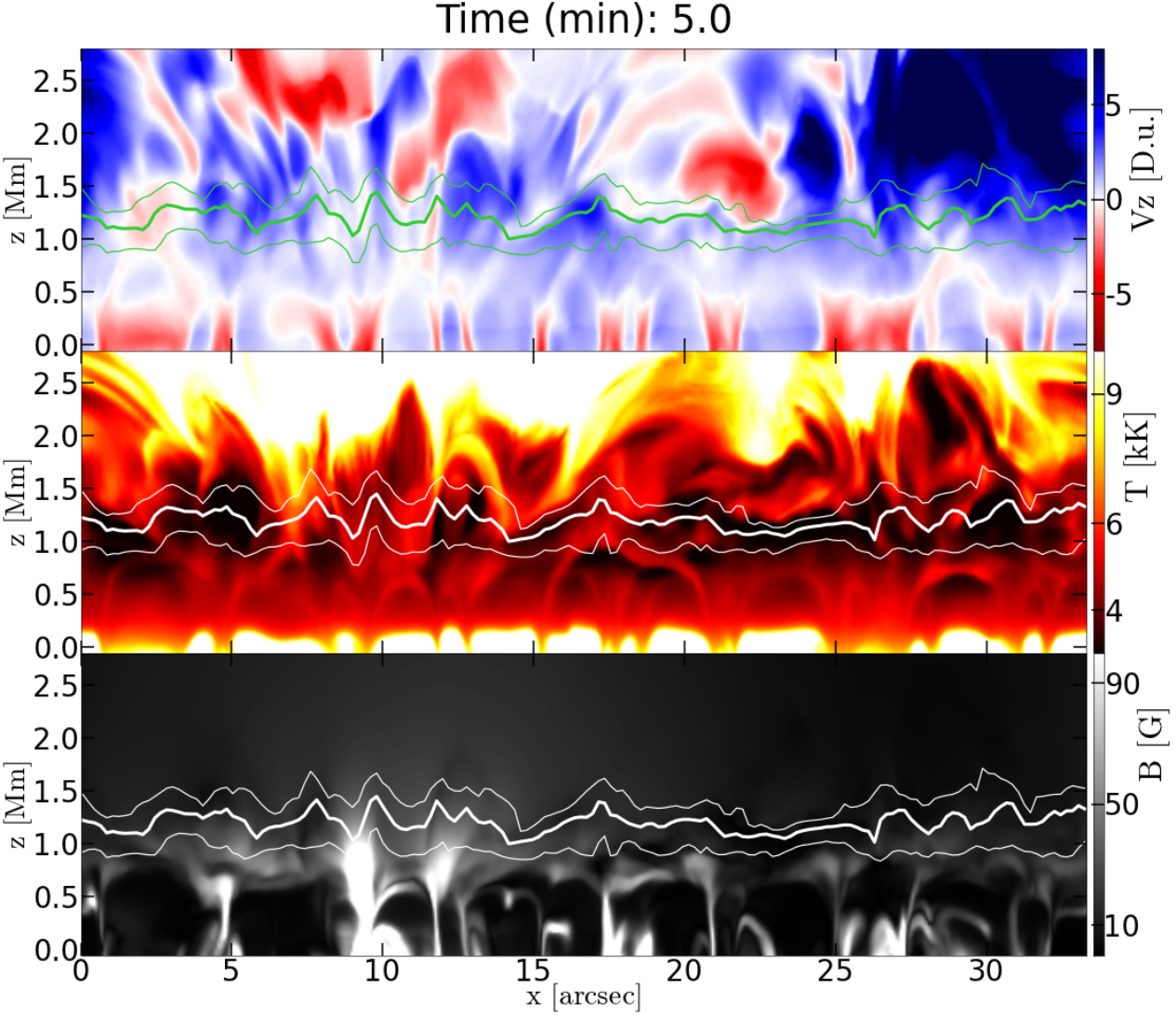}\\ 
\end{array}$
\caption{Left: magnetic field strength saturated in $100$ G at z=$0.7$
  Mm. Right: the formation heights of Ca {\sc i} $4227$ {\AA} line samples the low
  chromosphere above the small-scale magnetic canopy. The thin white lines mark
the formation region $\tau \in[0.1,10]$ for the wavelength of the
intensity minimum. The thicker white line marks $\tau=1$.}
\label{fig:fig1}
\end{figure*}

\subsection{The atmospheric model}
The input for our calculations is a time-dependent MHD
simulation computed with the
Bifrost code \citep{Gudiksen:2011aa} considering non-equillibrium hydrogen ionization. It
emulates a bipolar
magnetic structure with network properties and its quiet sun 
surroundings, having and average unsigned magnetic field strength of $48$ G in the model
photosphere. The spatial physical
domain covers a horizontal extension of $24 \times 24 \,\rm{Mm}^2$ with
a horizontal resolution of $48$ km and a vertical resolution of $19$
km in photosphere and chromosphere. The temporal evolution lasts $15$
minutes of solar time with a resolution of $10$ seconds. 
For more details see \cite{Carlsson:2016aa}.
Figure \ref{fig:fig1} shows the slit-like region of $\approx
0.^{\prime\prime}5\times 33^{\prime\prime}$ (it has a certain
width) that was selected in the models for our calculations.

\subsection{Calculation procedure}
We developed a pipeline of programs that processes data levels in independent steps. In the step $1$
 the MHD simulation (data level 1) is read and transformed for multidimensional visualization and plotting (data level 2). 
Having selected the region where the radiative
transfer is to be carried out, the inputs (data level 3) for the RH
1.5D code \citep{Pereira:2015aa,Uitenbroek:2001} are created. Such code is set to solve the NLTE ionization balance 
between Ca {\sc i} and
Ca {\sc ii} using a $20$-level atomic model that accounts for the lower
transitions of Ca {\sc i} and the ground level of Ca {\sc ii} with $19$
continuum transitions and $17$
line transitions. Photoionization and inelastic collisional excitations/de-excitations due to electrons are 
considered for all levels. The inclusion of Ca{\sc iii} is negligible for computing the 
Ca {\sc i} populations because the population of the former starts to be significant from 
heights above the upper chromosphere, while Ca {\sc i} forms entirely below the middle chromosphere.
 Thus, in the solar models considered, Ca {\sc ii} provides all the reservoir population.

The calculation of atomic populations with RH was done considering
partial redistribution \citep{Leenaarts:2012aa}.
Comparing the results in PRD and CRD we have seen that this 
affects (not dramatically) the atomic populations.  The
reason is simply that the NLTE mean radiation field is slightly
affected by the increased PRD emissivity. 
 
The atomic populations resulting for the levels of the $\lambda4227$
line and the MHD quantities are the input
(data level 4) for Handy$^\prime$ (HANle DYnamic Polarized Radiation In Moving
Envelopes). This code solves the non-LTE radiative transfer problem of
the second kind \citep[][sec.14.1; LL04 hereafter]{LL04} processing
each time-step independently
 and applying the 1.5D (or column-by-column) approximation to each
 pixel of the slit.
Thus, horizontal inhomogeneities and
 horizontal velocity gradients do not contribute to the non-local part
 of the problem (the radiation
 field). 
The non-LTE iteration provides the converged values of the 
 the components of the statistical density matrix, which accounts
for atomic populations, atomic polarization and quantum
coherences in magnetic energy sublevels. 
The emergent
radiative transfer performed from the 
converged atomic density matrix is fully realistic in forward-scattering (disk
center line of sight).
The local physics (collisions, Zeeman and Hanle effects) is properly treated. 
This approach allows to investigate
 the effect of vertical variations in the MHD quantities
 and provides of spatio-temporal
continuity to the results.

The calculations with polarization were done in the regime of complete redistribution
(CRD). This is justified because, though spectral PRD wings are increasingly generated 
for decreasing $\mu$ values, partial redistribution does not
affect the line core profiles at disk-center, meaning a range of
$\mu=[0.89,1]$ for the $\lambda4227$ line \citep[e.g.,][]{Dumont:1973aa,Anusha:2011aa}.
Our results point out that this is true in a more restricted disk
center area, roughly for $\mu\in[0.96,1]$, but
for smaller $\mu$ values the combination of
chromospheric velocities and photospheric velocity drifts with lack of
time resolution can make that the line core linear polarization (LP) and the near-core PRD
wings overlap in wavelength (see
Letter or Sec.\ref{sec:peakrings}).

In any case, the $\lambda4227$ line core is not blended or affected by the
weak spectral lines
forming in its proximities \citep{Lites:1974aa}.  
Continuum
polarization is generally very small and is also minimized in forward-scattering. 

\begin{figure}[h]
\centering%
\includegraphics[scale=0.55]{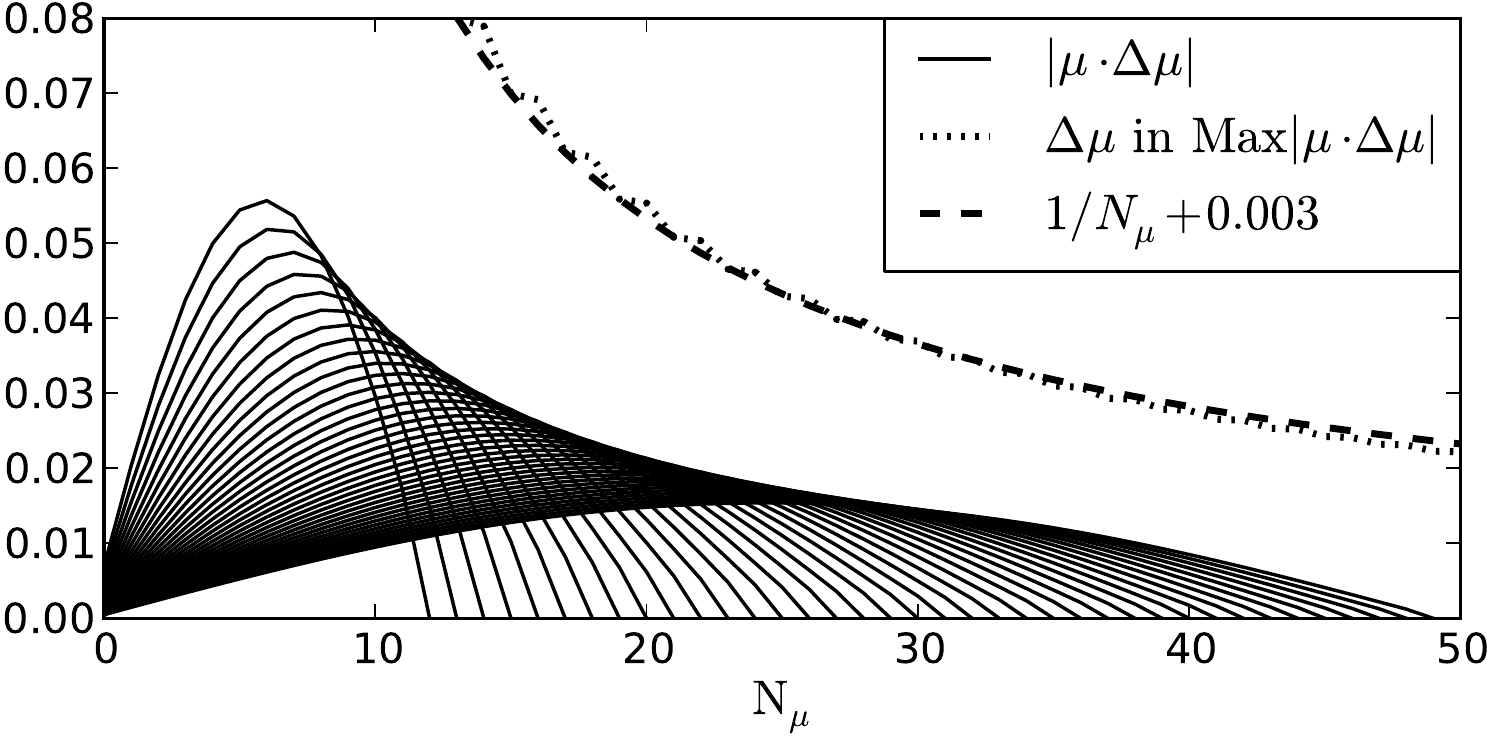}
\caption{In presence of large vertical
 velocities the error in the gaussian quadratures used to calculate the
  radiation field tensor is larger for those rays whose $\xi=|\mu \cdot \Delta
  \mu|$ is maximum (with $\mu=cos(\theta_{\rm{k}})$, and
  $\Delta\mu=\mu_{\rm{k+1}}-\mu_{\rm{k}}$, for a ray $k$). 
  Continuous curves above show this quantity for
 different quadrature grids with number of points labelled by their
 cut with the horizontal axis. In order to give a rule for the number of angular
 points $\mathrm{N_{\mu}}$ to be used, we take the $\Delta \mu$ at that 
 maximum $\xi$ as the limiting worst case. We find that 
 such $\Delta \mu$ is $\approx 1/\mathrm{N}_{\mu} + 0.003$ for any number of points.
 On the other hand, as $\Delta
 \mu$ must be $<\mathrm{1/V^{\mathrm{Max}}_z}$ ($\rm{
   V_z^{Max}}$ being the maximum Doppler velocity in the atmosphere), the
 recipe turns out to be 
 $\mathrm{N_{\mu}\gtrsim 2.1 \cdot V_z^{Max}}$. This rule is only
 applied when giving a number larger than a safe minimum 
 of 13 points per quadrant in inclination.}
\label{fig2}
\end{figure}

For this paper the pure Zeeman
effect (no atomic polarization, no quantum coherences)
in Q, U and V was calculated separately from the Hanle signals in
order to compare both contributions to the LP. Namely the equations solved here are: the
statistical equilibrium equations given by a suitable
combination of Eqs. (7.2.a) and Eq. (7.101) of LL04 under the
impact approximation and the
assumption of isotropy for depolarizing and inelastic collisions\citep{Lamb:1971aa} ; and the radiative
transfer equation for an instantaneously-stationary radiation field
with propagation matrix given by Ecs. 7.2.b (Hanle regime) and by
Ecs. (9.4), (9.7) and (9.10) (Zeeman regime) of LL04 neglecting
stimulated emission. 
The optical profiles entering in the propagation matrix
are calculated using 
a damping parameter that includes the dominant contributions of 
radiative ($A_{ul}=2.18 \cdot 10^8$) and
Van der Waals ($\gamma_{VdW}=1.7\cdot 10^{-8}$, $a_{VdW}=0.389$)
broadening \citep{Faurobert-Scholl:1992aa,Stenflo:1974aa}. 
The resolution of the wavelength and angular grids are
automatically set by adapting them to the level of kinematics
affecting the radiative transfer. When integrating in angle (e.g., to
obtain the radiation field
tensor) we use a gaussian quadrature whose
minimal number of points in inclination angles is defined by
the rule explained in Fig.\ref{fig2}.

The final step of the calculations tries to facilitate the analysis. 
It involves the characterization of
the Stokes signals and of the physical quantities related to
polarization at the region of formation of each 
wavelength. These metrics (data level 5) allow us to correlate
detailed quantities in multiple dimensions for understanding
possible patterns that can lead to better diagnosis in
time-dependent atmospheres. All data levels
are structured following NetCDF4 standards\footnote{http://www.unidata.ucar.edu/software/netcdf/}.
All our calculations were done for microturbulent velocity 
$\rm{v_{micro}}=0\, \rm{km \,s^{-1}}$ and $\rm{v_{micro}}=2\,\rm{km
  \,s^{-1}}$.  

\subsection{Minor modifications to RH 1.5D}
 In presence of
shock waves the
numerical convergence of some radiative transfer codes with
polarization is usually not
guaranteed because a mere Doppler shift can make the elements of the
propagation matrix of consecutive points in the optical path to differ
abruptly in amplitude at a given wavelength. Physically, this is
an obstacle specially in the low chromosphere. On one hand, because there the ratio
between the vertical velocity gradients and the thermal broadening is usually larger
than at other heights (by combination of shock waves and cool plasma
pockets). On the other hand, because in the low chromosphere the frequent meeting between upward shocks
and plasma falling back fast from previous waves produces larger velocity gradients. Numerically, the
problem is 
that some formal solvers of the radiative transfer equation become unstable
in such situations unless a very fine grid is used. Instead of modifying the formal
solver used by RH we have assured stability and
convergence modifying RH $1.5$D for \textit{redistributing}, when
necessary, the atmospheric grid points towards
those heights where certain proxies to opacity change more abruptly. Combining this
method\footnote{To be
  eventually published in a separate publication.} with an eventual 
better grid resolution all columns converge. 

Other minor modifications done to RH for developing this work
include: i) calculation of
 heights of formation for all wavelengths at additional optical depths in the transition of
 interest; ii) possibility of redefining the cutting atmospheric points using optical depth
 and hydrogen density thresholds respectively; iii) possibility of
 increasing the number of spatial grid points in run time.

\section{Results}\label{sec:results}

\subsection{Magnetic references in semi-empirical models.}\label{sec:hanle} 
\begin{figure}[h]
\centering%
\includegraphics[scale=0.6]{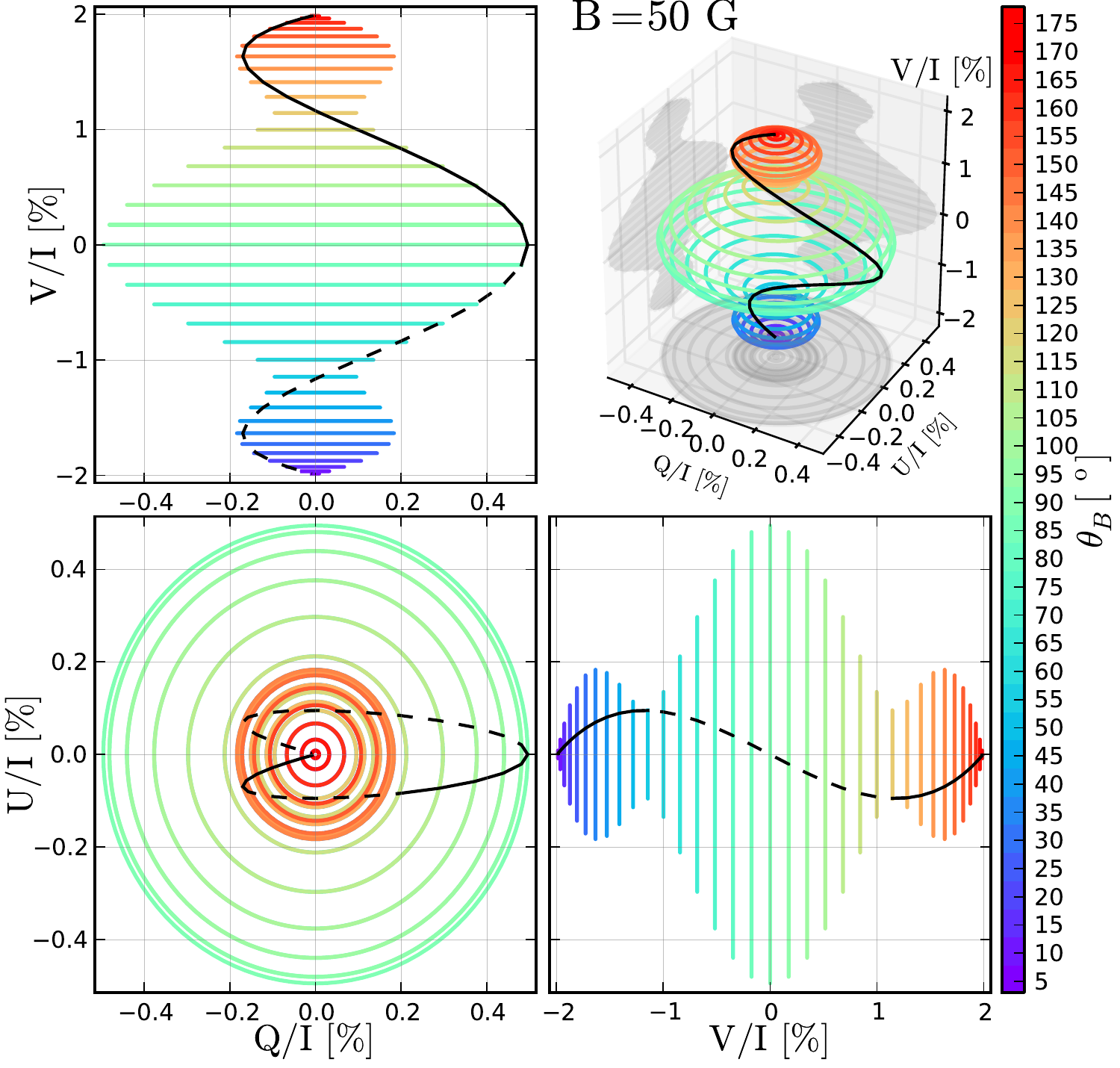}
\caption{Poincar{\'e } diagram for: Ca {\sc i} $\lambda4227$, $\mu=1$,
  FALC atmosphere, constant ad-hoc B$=50$ G. The black line connects
  all points with $\chi_B-\chi=90^{\rm o}$ ($\chi$ sets the direction
  of Q$>0$ in space). }
\label{fig:figpre3}
\end{figure}

Before considering MHD models we calculated the emergent
forward-scattering polarization in a FALC model \citep{Fontenla:1993}
with a constant ad-hoc
magnetic field. We did it for all
possible magnetic field azimuths and inclinations, and for strengths between $10$ and
$130$ G (Hanle saturation for Ca {\sc i} $\lambda4227$) every $10$ G. 
 Representing in the Poincar{\'e}
sphere (Q,U,V space) the amplitude of each
polarization profile for each case, we obtain an extension of a Hanle
diagram, what we call a Poincar{\'e} diagram(Fig. \ref{fig:figpre3}). This representation has
not been used before for characterizing the Hanle and Zeeman effects in a solar
atmosphere, but it seems quiet advantageous to this regard. The additional Stokes V dimension in Poincar{\'e} diagrams partially
breaks the degeneracy of polarization with magnetic field
orientation. This representation gives a more compact
and clear view of the limiting polarization values for a given spectral line
and line of sight. 

 We find that at disk center the
total LP of $\lambda 4227$ in semiempirical models is always in the
range $0.1 - 0.5\%$ for any magnetic field inclination $\theta_B\in
[17^{\circ},163^{\circ}]$ 
\begin{figure}[h!]
\centering$
\begin{array}{c} 
\includegraphics[scale=0.55]{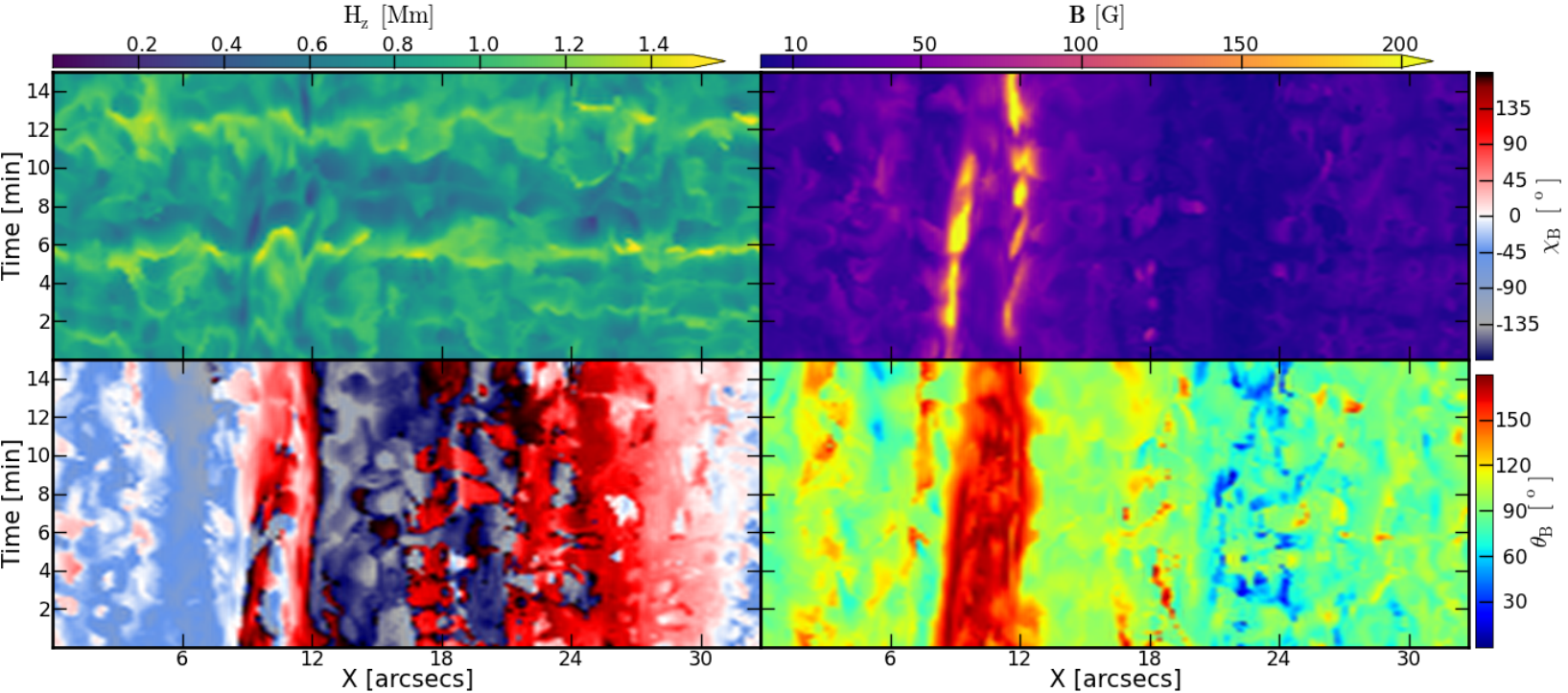}\\
\includegraphics[scale=0.55]{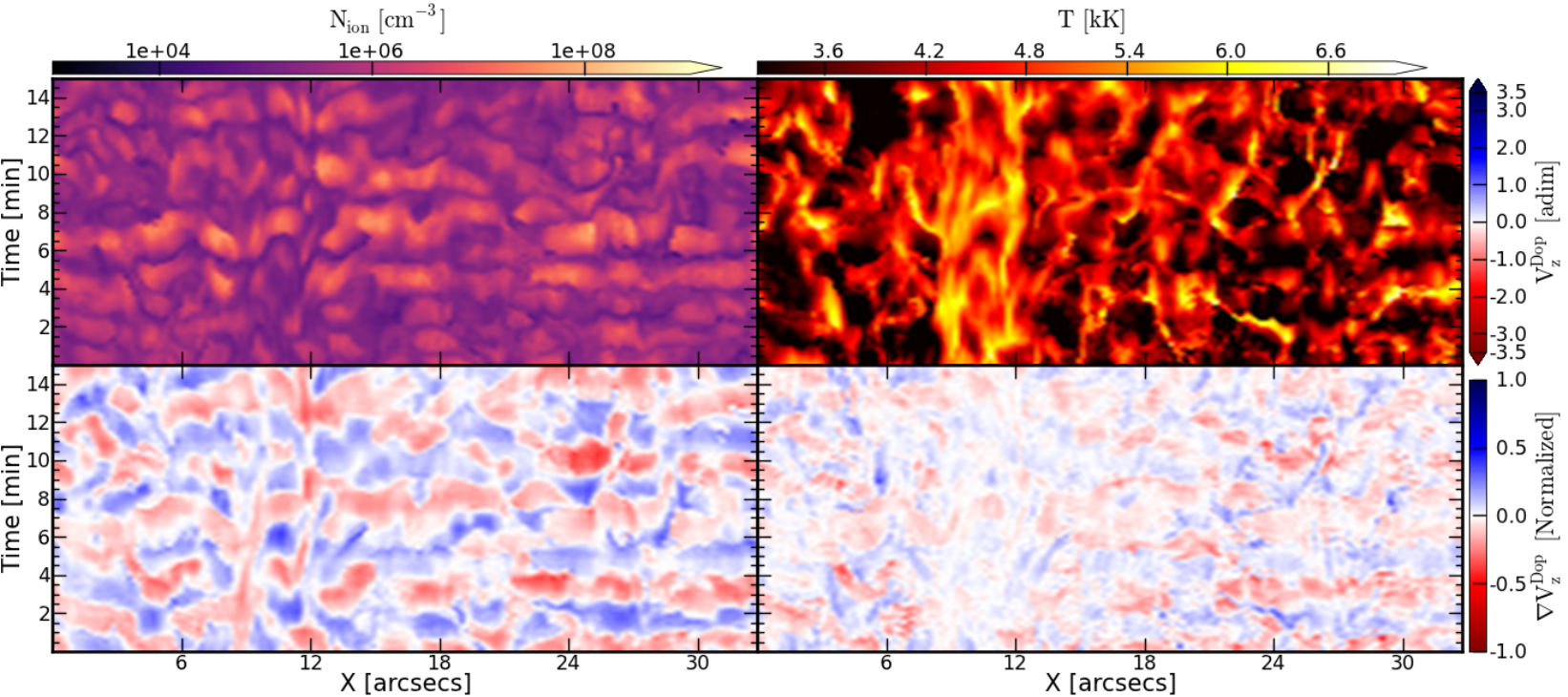}  \\
\end{array}$
\caption{MHD quantities at $\tau=1$ for
  the line-center wavelength in each timestep and length coordinate
  in the slit. Velocity shocks are here exposed by removing at each
  timestep the average offset velocity due to the $5$-minute
  oscillations (see its effect in the height of formation).The
  vertical velocity map is on the left.} \label{fig:figx3}
\end{figure}
and any azimuth if the magnetic field is lower than $50$ G. Most magnetic fields affecting the line core in the time-dependent
simulations are between this value and
$10$ G, hence very close to the optimal value that maximizes the Hanle
effect and that is given by the upper-level Hanle critical field of
$20-25$ G.

As shown by Fig. \ref{fig:figpre3}, the previous minimum of
LP=$0.1\%$, 
given by near-vertical magnetic
fields, is also produced at Van-Vleck inclinations:
$\theta_B=[54.73^{\circ},125.27^{\circ}]$. 
Therefore, predominantly
horizontal fields, with inclinations contained between those Van Vleck
angles, cannot make Q/I, U/I and V/I to be below
$0.1\%$ in large areas simultaneously. If this happens observing a line whose S/N
is expected to be good, we assume that is because collisions
and/or dynamic effects 
are cancelling the polarization. In particular, a total LP $<0.1\%$
over large solar areas suggests that dynamics might be producing sign cancellations
below the temporal resolution element (we will discuss this later). Here \textit{dynamics} 
also refers to situations with a time-varying magnetic field 
in the resolution element. For instance, our
simulations show that emerging cool
bubbles seem to make the
magnetic field inclination to oscillate between the horizontal and
near-Van-Vleck angles (see Fig. \ref{fig:figx3} around x$=18^{\prime\prime}$ and
x$=23^{\prime\prime}$). 

If the maximum values of the
measured LP are above the semi-empirical (static)
reference of $0.5\%$, it is necessary to assume that the amplifications introduced by macroscopic motions along
time are significant, occurring at certain wavelengths with enough persistence (spectral coherence) and/or strength during the exposure time.  

\subsection{Formation region}\label{sec:formation} 
Significant variations in the height of formation are normal in chromospheric
lines. In the models considered the region of formation
($0.1<\tau <10$) of the minimum line-core intensity of $\lambda4227$ oscillates between $0.7<z<1.5$ Mm, tending to
contain the coolest atmospheric patches located just above the
small-scale magnetic canopy (see Fig. 1, right panel). 
A first reason for this is that neutral Calcium density
peaks in cool volumes. A second one is that
in forward-scattering 
the height of formation is the lowest possible. Due to the proximity
to the small-scale magnetic canopy, the region of $\tau=1$ at the line
core is normally filled by 
near-horizontal magnetic fields at any time, which maximizes
the forward-scattering Hanle effect. Such magnetic canopy separates
photosphere and chromosphere (see Fig. 1) and seems to play a role in the 
heating of the chromosphere \citep[e.g., ][]{Goodman:1996aa}. Also in these layers the incipient shock waves start to act significantly in the line core. In semiempirical
models this
corresponds to heights between the temperature minimum and
the first temperature bump. A similar scenario is expected for 
chromospheric polarization signals of other neutral atomic
elements.
\subsection{Instantaneous polarization features}\label{sec:snapshot} 
The temporal variations of the
 synthetic polarization profiles along the slit are large. The LP has 
 almost-ubiquitous, sudden and conspicuous increments (in
absolute value) moving rapidly along the spatial direction.
An inspection of the instantaneous slit profiles in the
temporal serie \citep[e.g., in Letter, Fig. \ref{fig:fig4_5} in this paper, or Fig. 3
in][]{carlin:16b} reveals:
\begin{itemize}
\item the spatial exclusion (or complementarity) of linear and
  circular polarization due to their different sensitivities to
  transversal and longitudinal magnetic fields. 
An additional reason is that the formation region, though corrugated,
is roughly parallel to the surface, hence crossing
  suddenly the vertical magnetic lines emerging from magnetic
  patches. In observations at disk-center, sizable V/I and line-core LP are sometimes co-spatial. 

\item a weak (negligible) transversal Zeeman effect along the whole
  slit, though $\lambda4227$ forms in
  the low chromosphere. Hence, all relevant features in Q and
  U are Hanle polarization. As the field is relatively weak, the Zeeman profiles only have 
  $\sigma$ components. They usually enclose the spectral Hanle core, but sometimes
   they lay in it due to height-dependent longitudinal motions (see Fig.\ref{fig:fig4_5}). \textit{Each} $\sigma$
 component is narrow (even with microturbulence) and can be, each of them,
 antisymmetric,
 so having opposite signs in their small spectral width. When this happens the variable Doppler shifts existing
  during the time integration can easily weaken the final LP Zeeman amplitudes.

\begin{figure}[t!]
\centering
\includegraphics[scale=0.45]{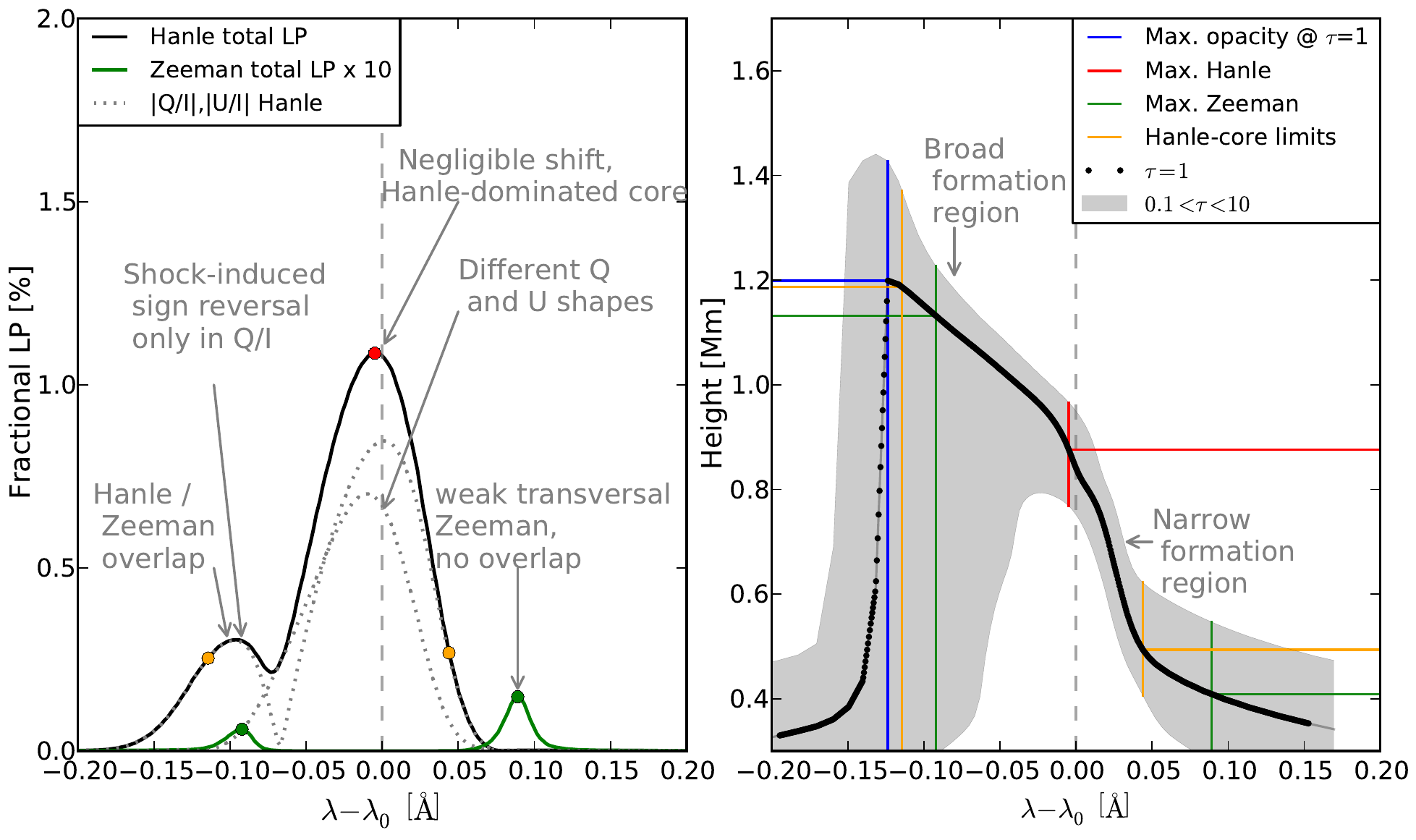}
\caption{Typical Hanle and Zeeman signals (left) and formation
  region (right) for one pixel and timestep. Total LP is defined
  as $\sqrt{Q^2+U^2}/I$. The Zeeman signals are multiplied by $10$.
  The yellow dots and lines mark $30\%$ of the
  Hanle maximum. Most of the LP features labelled in the
  figure change with time and pixel.}
\label{fig:fig4_5}
\end{figure}

\begin{figure*}[t!]
\centering$
\begin{array}{cc} 
\includegraphics[scale=0.7]{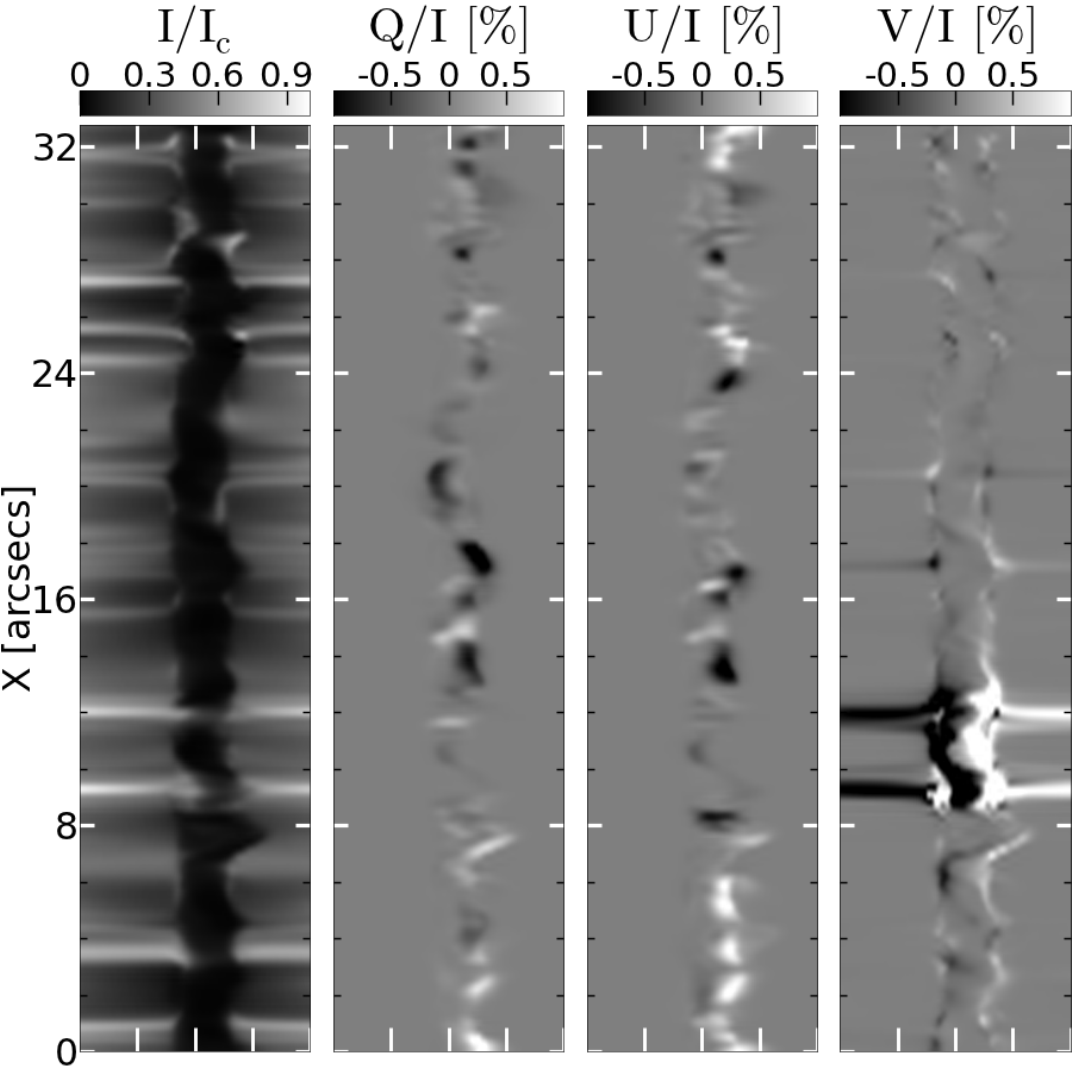}   & 
\includegraphics[scale=0.7]{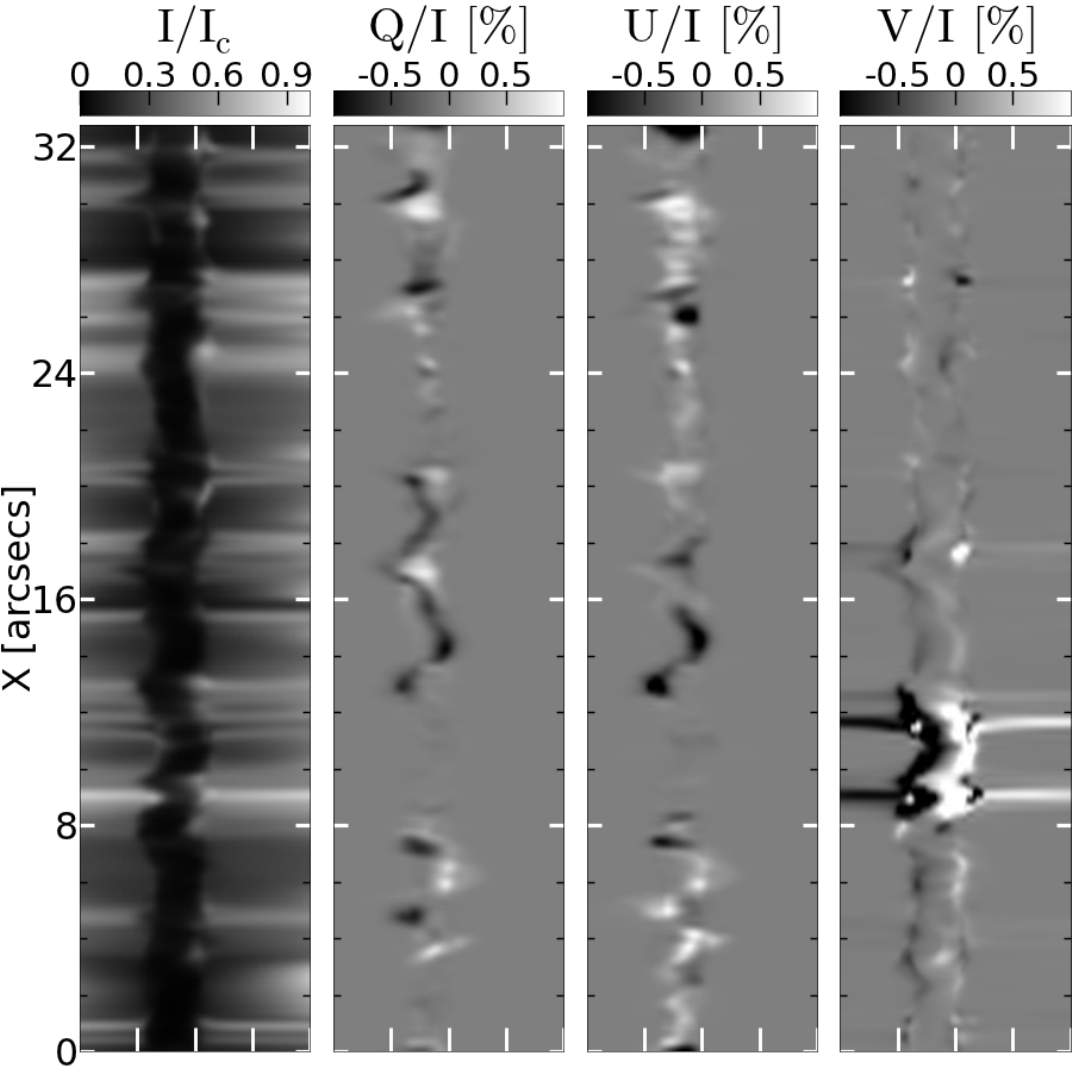}\\ 
\includegraphics[scale=0.7]{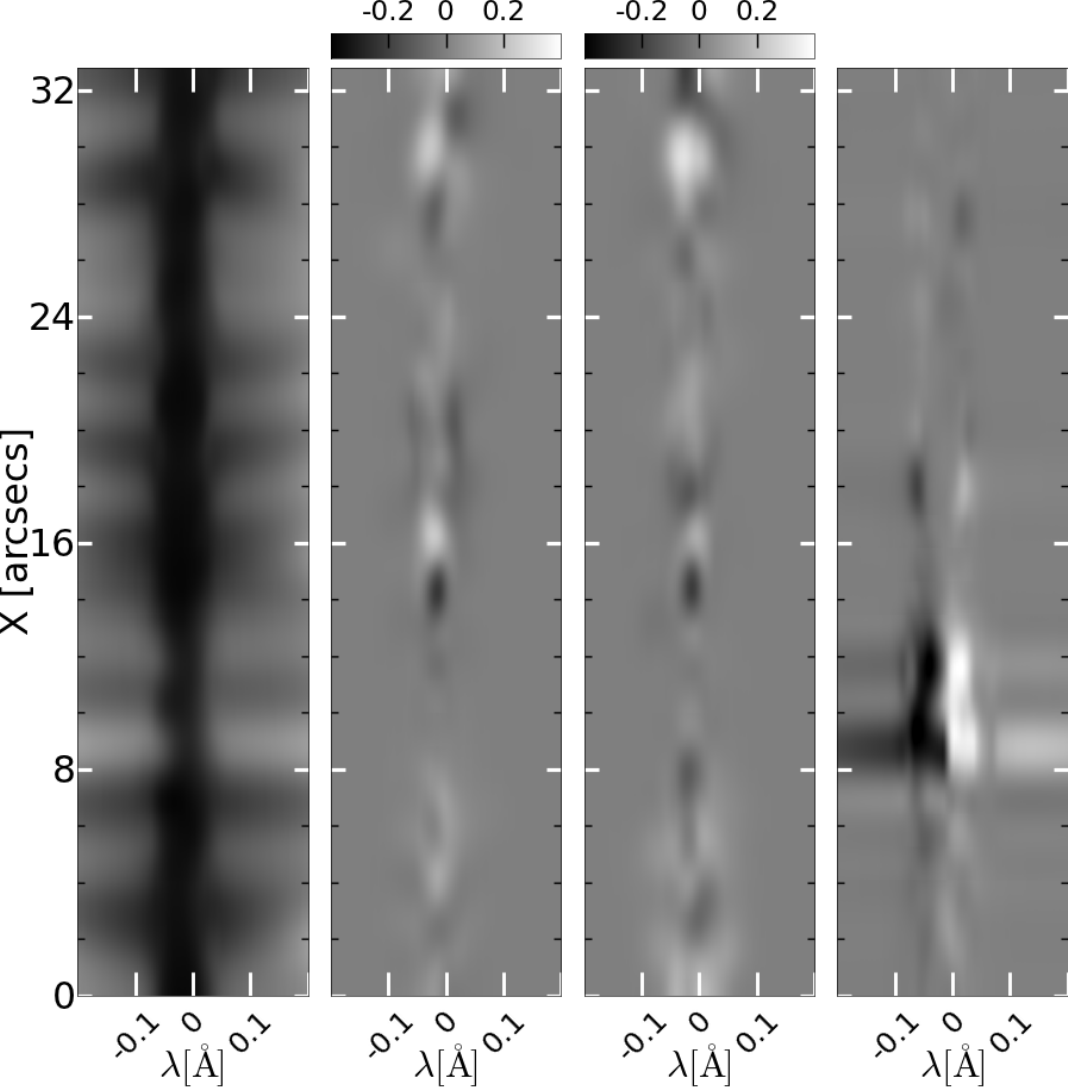}   & 
\includegraphics[scale=0.7]{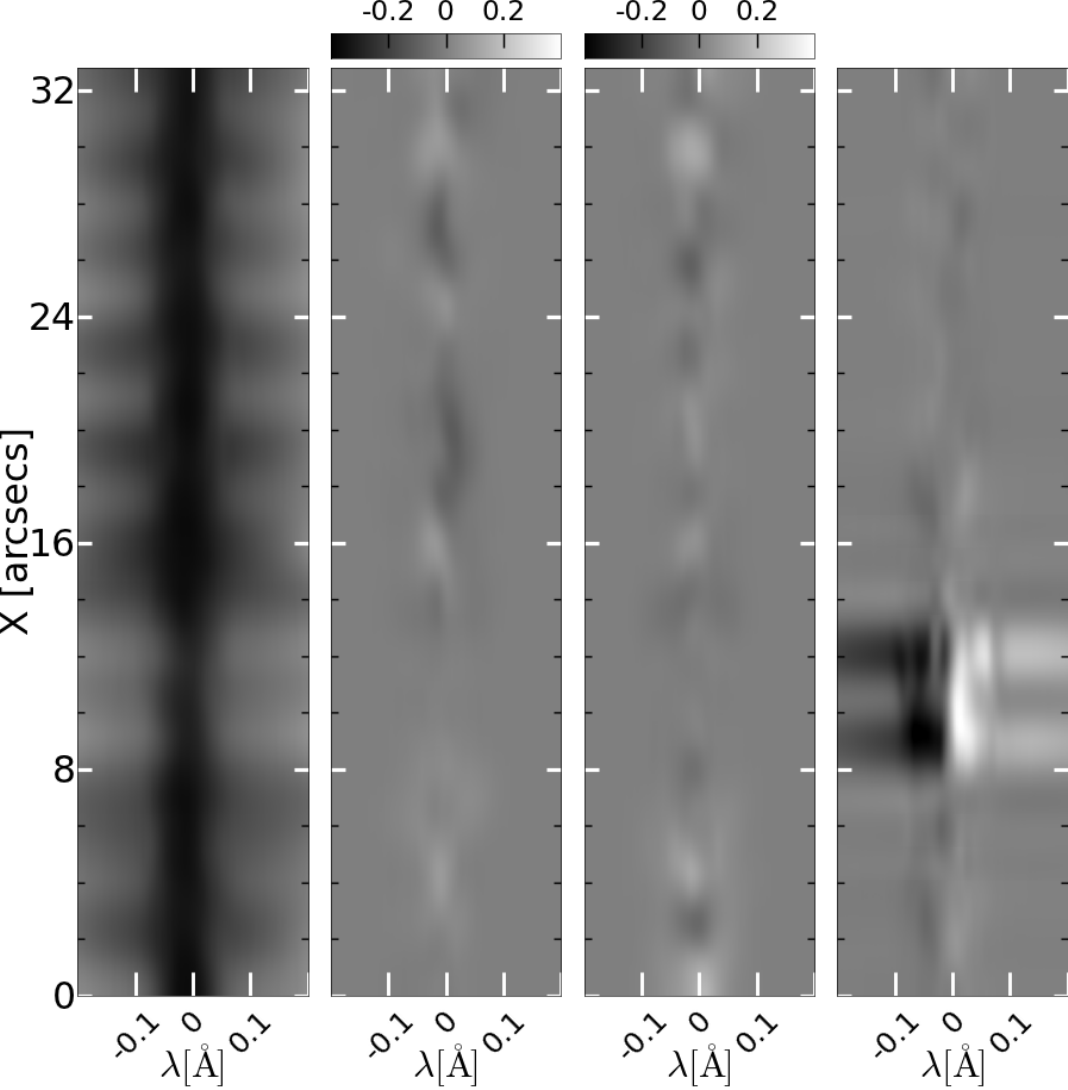}\\ 
\end{array}$
\caption{Slit profiles for $\rm{v_{micro}}=0\,\rm{km
    \,s^{-1}}$. Upper panels: instantaneous signals in minimum (left) and
  maximum (right, $150$ s later) of the supergranular
  convection. For calculating these panels we integrated
  each individual Stokes component in an area of $0.5$ (slit width) $\times
  \,0.2$ arcsec$^2$. Lower panels: same as before but integrating
  $1.4$ arcsec along the slit and 
  $5$ min (left panels) or $15$ min (right panels) in time
  . LP is saturated to $\pm 1\%$ or  $\pm 0.4\%$ as
  shown by the corresponding bars above each panel.
}
\label{fig:fig3}
\end{figure*}

\item the correlation between spectrally-broad strong circular
polarization, heating signatures in intensity and stronger vertical
magnetic fields. Spectrally-broad V/I profiles can be understood as a
 consequence of the weak-field approximation for Stokes V (being proportional
to the wavelength derivative of the intensity) when the intensity
profile has conspicuous peaks aside of the core and the longitudinal
magnetic field is strong enough.
At disk center, the formation region of the $\lambda4227$ near-core peaks in
intensity is already sub-chromospheric, hence the heatings creating
such intensity excesses are not due to compressing shock waves.
 
\item the dynamic modulation of Hanle signals by vertical
gradients of velocity. The general response of the
scattering LP signals to velocity gradients does not require
lower level alignment 
\citep[as did for Ca {\sc ii} IR triplet lines in][]{Carlin:2013aa} if the upper level can instead harbour it.

\item the strong instantaneous amplitudes of the forward-scattering
 $\lambda 4227$ polarization. They are larger than in
observations presumibly because are calculated with spatial and
temporal resolution, which minimizes cancellations. This should mean
that magnetic field diagnosis via
Hanle in the bulk of the chromosphere is not physically limited by
too weak signals at disk center.

\item the presence of antisymmetric LP profiles, without dominant
  sign (see Figs. 2 and 3 of Letter or
Fig. \ref{fig:fig3}). Hence, the Hanle core can in principle have any shape, and not necessarily a single line-core lobe as usually thought.
The origins of such antisymmetric LP profiles are: a variation of  
magnetic field azimuth along the line of sight (LOS); and/or
the modulation of the height-dependent radiation field anisotropy by
vertical velocity gradients.

\item the different instantaneous shapes of Q/I and U/I at a given same
  location. Due to the physical symmetry at disk center in $1.5$D,
  one would expect Q/I and U/I with similar (normalized) shapes.
This fails because the magnetic
field azimuth changes along $0.1<\tau_{\lambda}<10$, so that the height maximizing Q/I
(i.e., where $\rm \chi_B=0,\pm 90, 180$) in a
wavelength $\lambda$ 
is different than in U/I (peaking in $\rm \chi_B=\pm45,\pm 135$). Thus, the
magnetic field can narrow down the formation region of the
polarization to specific layers. This also means that the difference
in shape between both normalized profiles
quantifies the magnetic field \textit{azimuth gradient} along the LOS.  
\newline
\end{itemize}

\subsection{Polarization without microturbulence}\label{sec:res1}
Large scale photospheric oscillations introduce an offset Doppler
shift with a period of $5$ minutes that simultaneously affect all spectra along the slit. 
Comparing signals calculated without ad-hoc microturbulent
velocity (see Fig. \ref{fig:fig3}) in opposite instants in which the maximum
$5$-minutes-period Doppler shifts are maximum,
 it is seen that such shifts are larger than half the broadening of the LP
 profile. When the chromospheric Doppler shifts (e.g., due
 to shock waves) are added, the result is LP
 profiles with reduced overlapping along time (compare upper
 LP panels of Fig. \ref{fig:fig3}). This weakens and shapes the
 integrated LP signals due to their lack of
reinforcement during the exposure
 time. Thus, integrating $1.4^{\prime\prime}$ and several minutes
 ($5$ and $15$ min in lower left and right panels, respectively) we find synthetic LP
 signals that are weaker than those in the observations
 of Fig. \ref{fig:fig_obs}. The maximum absolute LP value in the $15$-min integration is
  $0.11\%$, which is four times weaker than in the observations and three times
  weaker than in the calculations with $\rm{v_{micro}}=2\,\rm{km
    \,s^{-1}}$. Let us note that a distinctive point of the observed scattering
 polarization is to have significant amplitudes after long
 exposures, hence it is important to emulate this.
Furthermore, without microturbulence all the
Stokes components are too narrow. 

\subsection{Polarization with microturbulence}\label{sec:res1b}
\begin{figure}[h!]
\centering%
\includegraphics[scale=0.65]{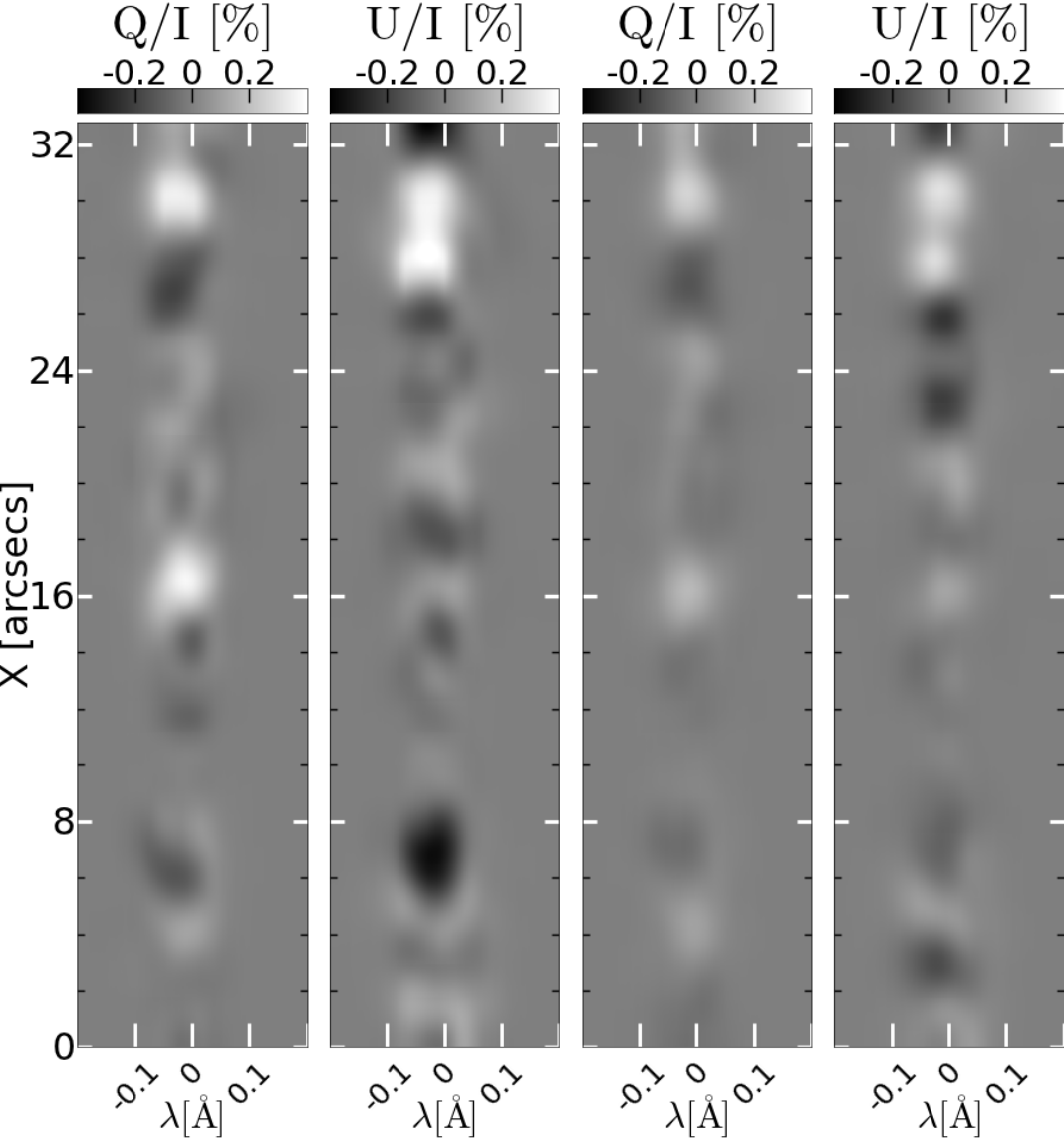}
\caption{Unresolved LP with $\rm{v_{micro}}=2\,\rm{km
    \,s^{-1}}$ and spatial resolution of $1.4$ arcsecs. First two panels:
  LP after 5-min integration. Last two panels:
  LP after 15-min integration. The action of microturbulence allows to
have measurable amplitudes after long exposures.}
\label{fig:figx1}
\end{figure}

The observational constraints in broadening and amplitudes mentioned before are reasonably
achieved when $\rm{v_{micro}}=2\,\rm{km \,s^{-1}}$ (see
Fig. \ref{fig:figx1}).  The significant improvement provided by the microturbulence 
points out that the lower chromosphere of the models is too
cool (particularly around the coolest locations, where this line tends to form).
 The agreement in morphology is
now also remarkable. A chain of near-symmetric LP rings\footnote{These
 LP features resemble rings in the wavelength-space plane because the core Hanle signal is
surrounded in space and wavelength by nearly symmetric peaks of
opposite sign.} appear after $5$ min of integration, when the maximum integrated
amplitudes coincide with the observed ones. Such an agreement between the left two panels of
Fig. \ref{fig:figx1} and the observations results of 
adding up hundreds
of very different instantaneous Stokes profiles (those in a bin of
$1.4^{\prime\prime}$ and $5$ minutes). 

An equivalent example for
longer exposure times ($15$ min)
is given by the second pair of panels in the same figure. The maximum
amplitudes are yet close to the observed ones, as required
in Sec. \ref{sec:res1}.
 Additionally, there is attenuation over significant
 extensions of the slits, with amplitudes well below the
 minimum reference defined in semiempirical models in
 Sec. \ref{sec:hanle}.  This can help to explain the ``noise pools'' found in 
observations of scattering polarization. For instance, the very
 small (noisy) V/I, Q/I \textit{and} U/I over
the lowest part
of the slit in the observation of Fig. \ref{fig:fig_obs} is surely not produced by a particular
magnetic field orientation\footnote{As shown in Sec. \ref{sec:results}
  with Poincar{\'e}
  diagrams, a Van-Vleck magnetic field inclination over the whole area and during the whole
  exposure time is quiet unlikely.} or by collisional processes,
which should act in other close IN regions as well. Hence, dynamics
might be the reason.
Middle panels of
Fig. \ref{fig:figx2} shows how dynamics and time integration attenuate
up to a factor $20$ our synthetic observations in relation
to the maximum time-resolved signal in the same panels. 
Comparing with other locations of the slit with less attenuation we found that strong
reductions are produced when the temporal evolution of the
signals loose periodicity after a certain time, such that the
 pattern is not reinforced over long temporal scales. For instance, this happens where LP rings appear.

As in the observations, the V/I profile with microturbulence has a
 central spectral gap with near-zero amplitude when magnetic fields
 are vertical. The maximum V/I amplitudes 
suggests that the temporal scales are correct for the quantities and
heights creating
circular polarization. Averaging the two
peaks of the profiles we find $[2.16-(-1.85)]/2\approx2 \%$ and
$[0.84-(-0.82)]/2\approx 0.83
 \%$ (for the first and fifth strong V/I patch in Fig. \ref{fig:fig3}) versus $[0.96-(-0.84)]/2= 0.9
 \%$ (for our $15$-min-integration synthetic  signals)
This agreement is remarkable because 
the instantaneous V/I profiles are completely different than the integrated
ones (see lowest rightmost panel in Fig. \ref{fig:figx2}), which would
apply to real observations though being usually not considered
in the literature using the longitudinal Zeeman effect. This implies
that longitudinal magnetic fields inferred from quiet sun observations without
temporal resolution might be significantly
weaker than the true ones. 

As the calculations with $\rm{v_{micro}}=2\,\rm{km \,s^{-1}}$ seem to
represent better the sun, only this case is considered in the following.

\subsection{Emulating a longer slit}\label{sec:rings2}
\begin{figure}[h!]
\centering
\includegraphics[scale=0.65]{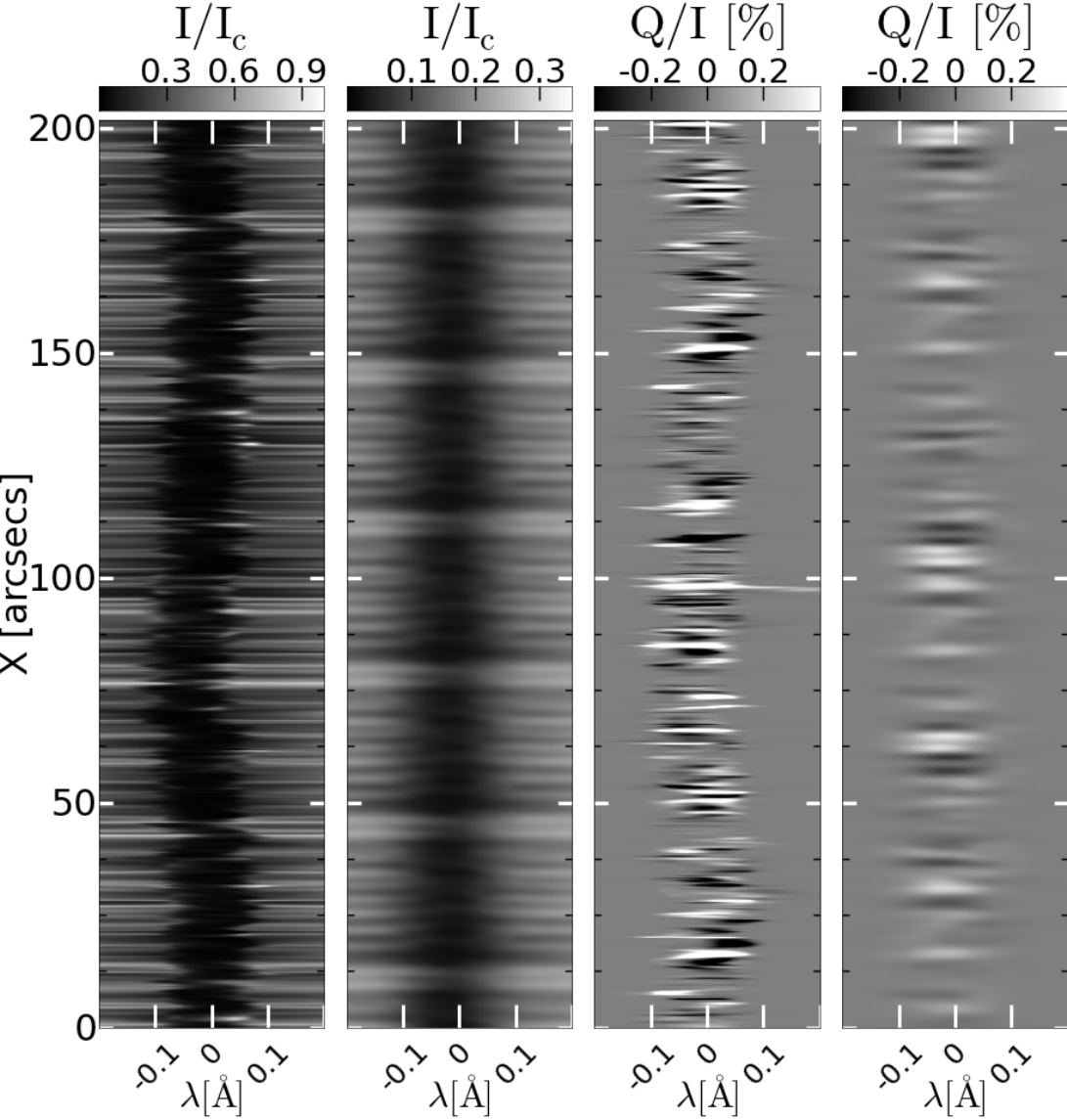}
\caption{Spatial pile of synthetic profiles filling the length of the observational
  slit. First two panels: Intensity profiles with full spatio-temporal
  resolution and after integrating $15$ min. Last two panels: same for
  Q/I. Actually,  to avoid repetition the instantaneous panels were
  composed piling several slits of
  different timesteps while the averaged slits where obtained by mere
  repetition along space of the average profiles. Similarly, the average Q/I panel is a
  pile of Q/I and U/I average slits.} 
\label{fig:figx4}
\end{figure}

By repeating blocks of intensity and LP profiles (Q/I and U/I indistinctly) along space, Fig.
\ref{fig:figx4} emulates a slit with the same length as in our
observations. Second panel shows the synthetic intensity integrated
over the $15$ minutes. Stokes $I/I_c$ has intensity variations
(strips) along the spatial direction whose small scales are
compatible with the observed $30$-min average
intensity in Fig. \ref{fig:fig_obs}. In this latter figure one has to
zoom a bit to see the less-contrasted strips (obtaining synthetic
intensity strips with a similar low contrast is just a matter of
adding the effect of the stray light in the telescope). 
The intensity strips are more evident toward the wings and seem to be due to sound
waves altering temperature. In the
simulations, they have more variability and contrast in
weakly magnetized patches of the IN because there the plasma
alternate periodically between lower and larger temperatures, which
increases emissivity and modulates the damping parameter shaping the
near-core. 
The synthetic intensity strips tend to disappear with stronger magnetic field, which correlates with reduced temperature and
 reduced vertical gradients of velocity. This suggests a relation between
 more active short-scale dynamics and LP rings.
Zooming in the intensity in Fig. \ref{fig:fig_obs}, we effectively have
the impression that there is a certain correlation between
the strips and the presence of LP rings in Q/I or U/I. 

On the other hand, right-most panel of \ref{fig:figx4} shows that the synthetic LP (Q/I and
U/I repeated in space) integrated over the $15$-min serie is spatially
different than in the $30$-min observations in Fig. \ref{fig:fig_obs}. 

For shorter integration times, the spatial lengths of the synthetic LP rings are at
least a factor 4 smaller than in observations.  The integration time
needed for reproducing their line-core amplitudes is also
significantly smaller ($5$-min) than the $30$-min of the
observations. In the Letter we tried to explain these differences in
terms of: i) the combination of Hanle effect, kinematics, and lack of
resolution, which are effects contained in our simulations; ii)
transversal Zeeman effect, which shows negligible contribution in our simulations; and iii) partial redistribution effects (PRD), not considered
by our calculations because in forward-scattering it plays no role,
but possibly affecting the observations that were not taken
exactly at $\mu=1$.
These factors behave differently in the two parts of the LP rings 
(the core and the near-core sign reversals surrounding it), hence 
they contribute differently to explain both parts.
We will now present additional information 
that can help to clarify the reason of the
differences in scale, starting by the contribution of the line core.

\subsection{On the core of the linear polarization rings}\label{sec:rings}

The length of the LP rings is set by the spatial change of sign in the very
line core. If the change was more frequent in space,
the number of rings in the slit would be larger and their scale shorter.
The maximum possible amplitudes of the synthetic and solar LP rings can be
 large (compared to the maximum reference values of
semi-empirical models, see Sec. \ref{sec:hanle}) and
occurs at line center, not in the sign reversals. 

In observations, the LP
rings are interconnected by single-lobe signals (forming a chain of rings in IN areas). 
Similar structures appear
 in our simulations close to the intermittent
 emergence of photospheric magnetic elements that are associated with small-scale 
oscillating chromospheric fields. Figure \ref{fig:figx3} shows this at
the level of the chromosphere. There, a regular emergence 
of cool plasma bubbles (see temperature panel around x$=17^{\prime\prime}$
and x$=22^{\prime\prime}$ between $0$ and $8$ min)
develops into more periodic chromospheric shocks at stable locations. 
When seen in space-time diagrams, the temperature in these locations show post-shock
rarefaction volumes, i.e. cool bubbles, that are periodic and look more rounded.
The magnetic field inclination, azimuth and
strength oscillate at such positions.
The inclination oscillates between horizontal and near-Van-Vleck angles exposing
opposite polarities for x$=17^{\prime\prime}$
and x$=22^{\prime\prime}$ in Fig. \ref{fig:figx3}. In between those
locations, the chromospheric magnetic field is always close
to horizontal. The single-lobe signals linking the LP rings vary in location around such middle
locations after integrating in time (Fig. \ref{fig:figx1}). 
The LP rings disappear after those $8$
minutes, when the repetitive pattern produced by the waves ends. 

\begin{figure*}[t!]
\centering%
\includegraphics[scale=0.65]{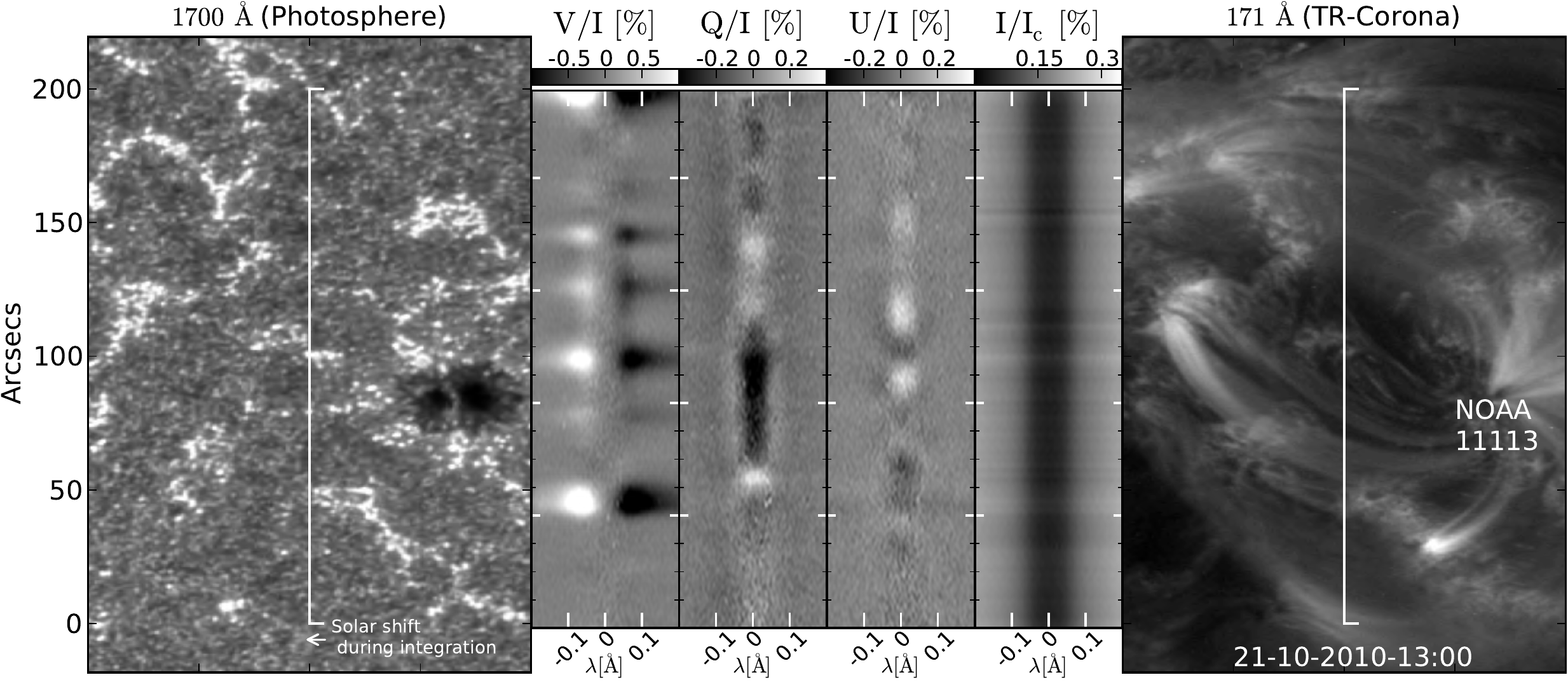}
\caption{Zimpol@IRSOL observation in $\mu=0.94$ (middle panels) and corresponding
  SDO images (lateral panels). The polarization colorbars are saturated
to $\pm1\%$ and $\pm0.4\%$. The strongest V/I signals are always associated to the network
 while the weaker ones correspond to
the weak photospheric magnetic
elements inside the IN. Zooming the intensity one can see that there is a
 certain correlation between the strength of the LP rings and shorter-scale
spatial intensity variations
of higher contrast. }
\label{fig:fig_obs}
\end{figure*}

\begin{figure}[h!]
\centering$
\begin{array}{c} 
\includegraphics[scale=0.45]{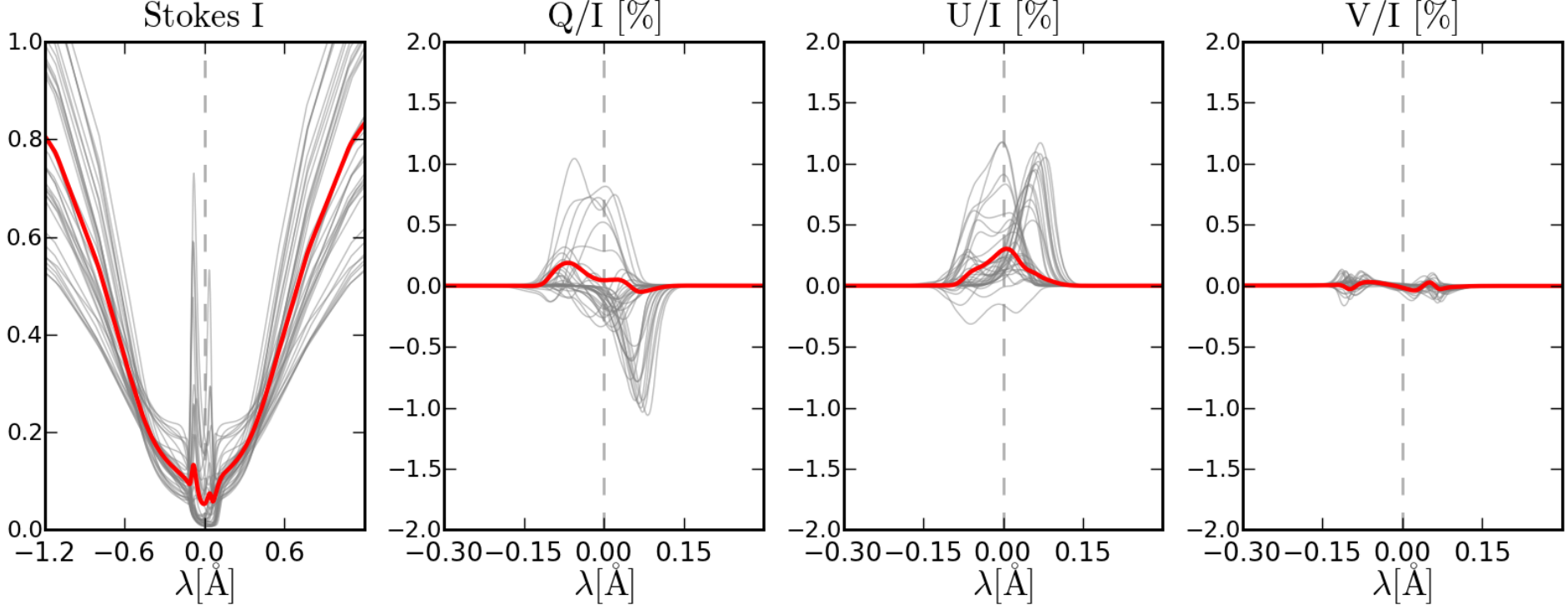}  \\ 
\includegraphics[scale=0.45]{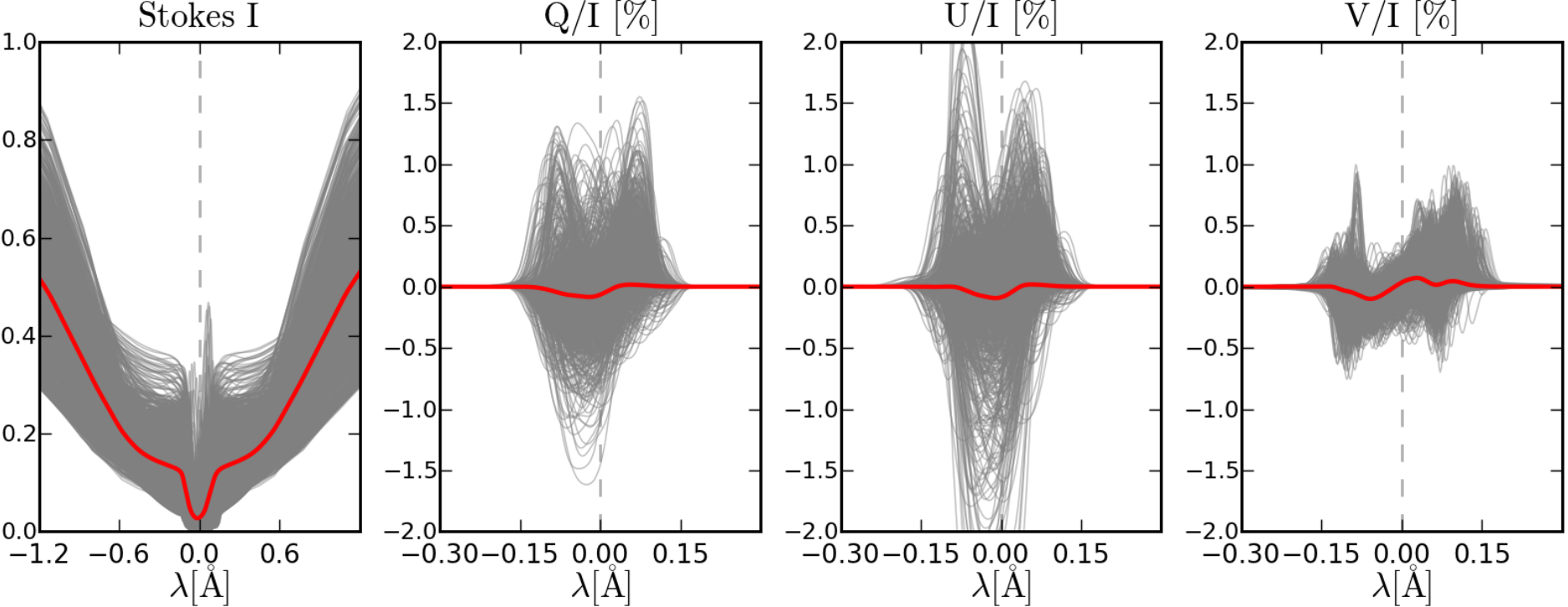}\\ 
\includegraphics[scale=0.45]{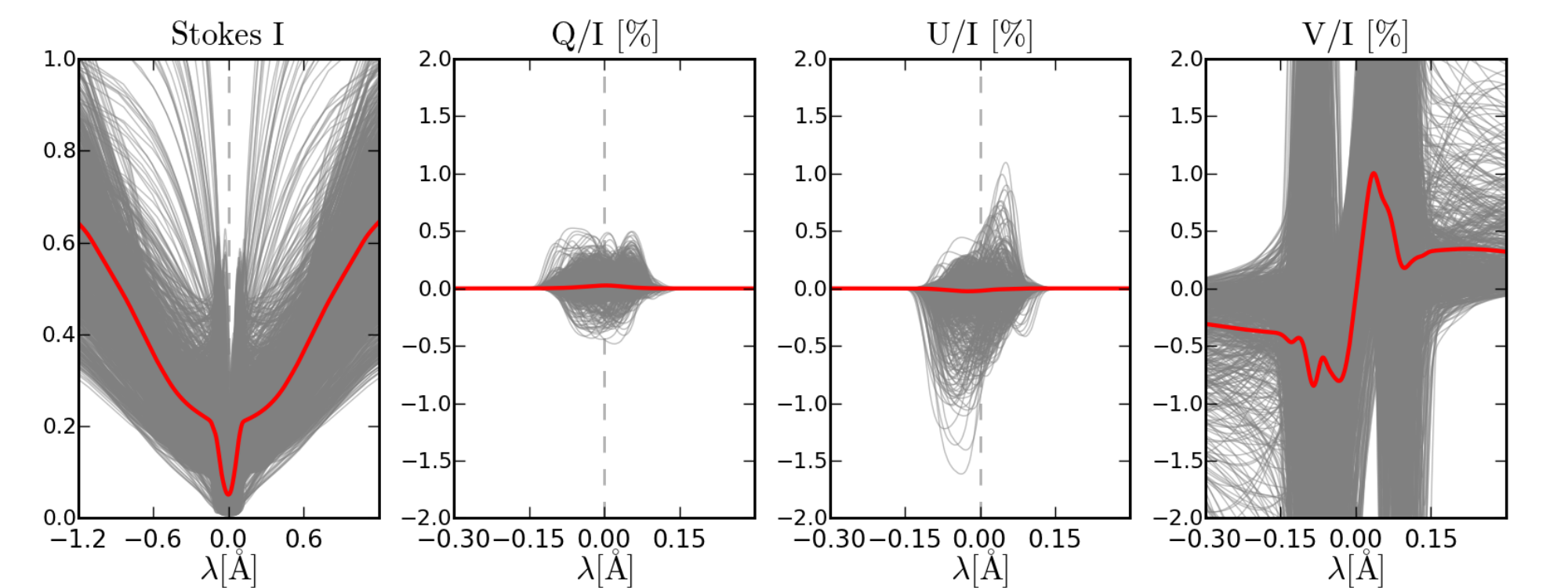}  \\ 
\end{array}$
\caption{Effect of short and long integrations with
  $\rm{v_{micro}}=2\,\rm{km \,s^{-1}}$. In gray: instantaneous Stokes
  profiles in a given spatio-temporal resolution element. In red: corresponding result
  of integrating each Stokes profiles in that bin and build the
  fractional quantities. Upper panels: $\Delta x=1.4"$, $\Delta t=20$ s.
Middle panels: $\Delta x=1.4"$, $\Delta t=15$ min in a pixel with
predominantly horizontal magnetic field in the low chromosphere during
the integration. Lower
panels: same than before but in a pixel with more vertical and
 intense magnetic field. } 
\label{fig:figx2}
\end{figure}

This picture presents some agreements with the observations. 
To have a glimpse of what happens below the temporal resolution element of the ZIMPOL
observations \citep{Ramelli:2010aa}, we have inspected the corresponding time
evolution of the SDO photospheric images (see one snapshot in
Fig. \ref{fig:fig_obs}). We find that the link between adjacent LP
rings lays between groups of photospheric magnetic elements that appear and disappear around the slit during the exposure time. This leaves a faint
residue in the integrated chromospheric V/I signals: see in Fig. \ref{fig:fig_obs} the two weak
and blurred internetwork V/I signals with opposite polarities
around $\rm x\approx 170$
arcsecs. This residual circular polarization is faint because the
emergent magnetic elements, being intermittent, are not there during the whole exposure
time. We think this because in our simulations the chromospheric V/I
trace of similar weak magnetic elements fade away easily with exposure
time too. Thus, we deduce that these small magnetic spots create field loops that reaches
the low chromosphere horizontally, changing the polarity of V/I and 
the LP sign of the Hanle core
 there where the two LP rings join. 

\subsection{On the sign-reversals of the linear polarization}\label{sec:peakrings}
The origin of the sign reversal conforming the outer part of the LP rings is not clear because
they lay in intermediate wavelengths where
all plausible physical effects seem possible.

A transversal-Zeeman scenario might seem reasonable
for explaining them because significant sigma
Zeeman components
can be produced
by a magnetic field 
increasing rapidly downwards from the formation heights of the Hanle core. 
 But several things point out the contrary. First, the observed LP rings have
  sizes compatible with the internetwork (IN) patches ( $\approx
  20-100$ arcsecs). This would imply large-scale magnetic
  fields acting in the IN at the level of the short-scale canopy. Namely,
this would mean that large-scale organized field lines (rooted in the
network, we presume) with significant chromospheric strengths
($\gtrsim 100$ G), are overlaying the
IN magnetic canopy after their expansion from the network photosphere to the
low IN chromosphere. This is incompatible with the current view of an
IN permeated by weak short-scale fields and with the simulations. 
In the simulation, the slit is located very close
to a network-like patch (see Fig. \ref{fig:fig1}) but, even so, the
transversal Zeeman signals are
roughly an order of magnitude weaker than Hanle. Furthermore, as explained
in Sec. \ref{sec:snapshot}, larger time
integration easily weaken the sigma components even more. Finally, the presence of symmetric sign reversals aside the core in the
observations is sometimes 
uncorrelated with the kind of magnetic structures (network or IN)
generating it. Hence, the transversal Zeeman effect can be ruled out.

Note that just a sign reversal in the height
variation of the radiation
field anisotropy cannot explain the two observed near-symmetric sign
reversals aside the LP core. As we expand now, there are several
reasons, related to how chromospheric motions modify the anisotropy and its impact in the
polarization. On one hand, there are
effects controlling the sign reversals \textit{in the resolved LP
  profiles}: first, the modulation of the anisotropy exerted by 
vertical velocity gradients can make the LP profiles 
antisymmetric in low-chromosphere spectral lines
(second-row right
panel of Fig.7 of Carlin et al, 2012.); second, the negative part of the anisotropy tends to disappears with increasing vertical
gradients of the source function, e.g. with
temperature gradients or in shock waves \citep[Fig. 4 of
][]{Trujillo-Bueno:2001aa,Carlin:2013aa}; third, when the anisotropy stratification has both negative and positive regions, the
dominance of each part in the LP profiles
varies significantly due the compression/expansion of the formation region
around shock fronts (Fig.5.3 of Carlin et al
2013); fourth, the microturbulent velocity tends to 
``erase'' the near-core sign reversals of LP; and fifth, the Hanle
effect of a magnetic field azimuth varying with height can also
produce a sign reversal in LP (see last point in
Sec. \ref{sec:snapshot}), at least around disk center,
where the solar-limb polarization offset is weak.
The net combination of these effects varies with the spectral line and
during the shocks emergence, but the theoretical trend in the chromospheric lines
that we have studied is \textit{to destroy the sign reversals in resolved LP profiles}. Thus, while
anti-symmetries (only
one sign reversal in the LP core) are possible and
assymmetries are everywhere in our simulations, it is
very rare to get simultaneous
sign reversals in both sides of the core. However, if on top of that
we analyze the \textit{unresolved} LP profiles resulting of such situation, we find
that dynamics can mimic quasi-symmetric sign reversals due to the combination of
instantaneous anisotropy-driven modulations of Hanle signals that are
quasi-sychronized with Doppler
shifts and integrated in space and time \citep{Carlin:2016aa}. The
middle panels of Fig.9 show how the instantaneous LP profiles (in gray)
tend frequently to group in a
sort of Zeeman $\pi-\sigma$ configuration.  Hence, the same
feature that was deleted by motions in resolved signals is recreated
in the integrated LP when time evolution is included. We refer to this whole
situation as ``dynamic Hanle'' scenario.

\begin{figure}[h!]
\centering
\includegraphics[scale=0.39]{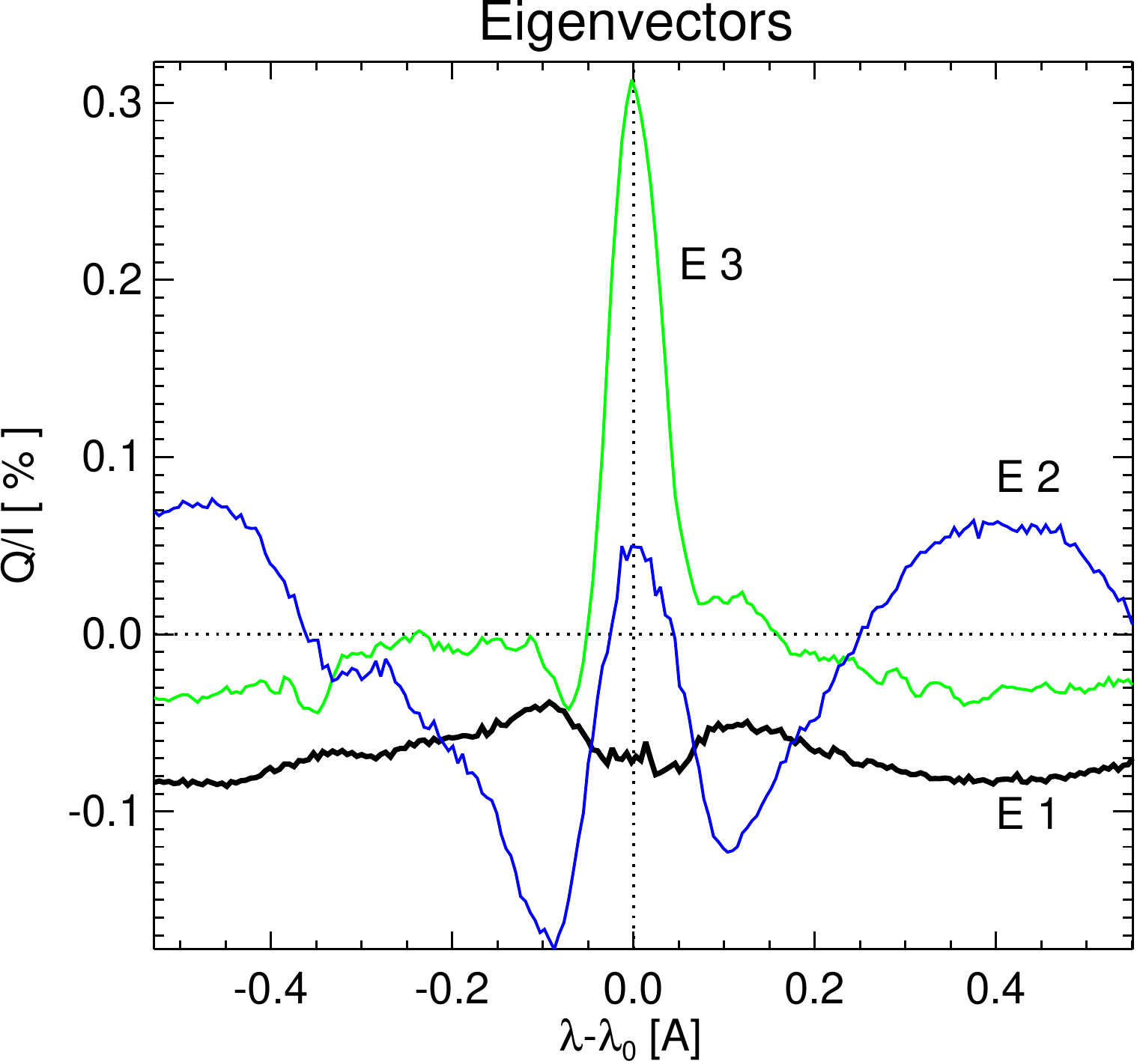}
\caption{PCA eigenvectors for Q/I profiles taken in various LOS's
  $\in \mu=[0.9,0.96]$. Note that the PRD minima (blue) is different
  than the Hanle minima (green). In Fig.(\ref{fig:pca2}) (Appendix),
  this eigenprofiles are used to reconstruct some observed profiles.}
\label{fig:pca}
\end{figure}

Thus, we are left with two explanations for the symmetric sign
reversals: dynamic Hanle and PRD. We try now to 
discriminate their relative influence by paying attention to the different location of the
 PRD and Hanle peaks. To show this we have applied Principal
Component Analysis \citep[PCA;
e.g.,][]{Rees:2000,Skumanich:2002,Martinez-Gonzalez:2008a} 
to observations of
Ca {\sc i} $\lambda4227$ done with ZIMPOL at IRSOL. We considered 
$450$ Q/I profiles in several lines of sight ($\mu\in[0.9,0.94]$)
including profiles with PRD effects around the core. Thus, we obtained the
first three PCA eigenvectors (eigenprofiles, see
Fig. \ref{fig:pca}) representing $95\%$ of the variance of the
data. They are ordered (E1,E2,E3) by the amount of statistical variance
that each one explains (size of the projections of the observed dataset in each eigenvector). 
The amplitudes of the
eigenprofiles are unimportant, only
shape matters. 

It is known that the first PCA eigenvector (E1) typically
represents the average of the data \citep{Skumanich:2002}, in our case mostly
affected by instrumental offsets, hence 
unimportant. In our analysis the
second eigenvector (E2) is capturing the PRD wings and separating them from
the line CRD core, which is represented by the third eigenvector (E3)
and dominated by Hanle and dynamics. 
PCA allows this separation because the contributions of
  dynamic Hanle and PRD are maximized at layers that 
  behave very different (chromosphere and sub-chromospheric
  layers). As this happens consistently in all pixels, both physical mechanisms produce statistically uncorrelated changes in the observed profiles, so that PCA get to separate them in
 eigenvectors.
The PRD
wings of E2 have a far-wing maxima, and also a near-core minima that we
 assign to the sign reversals discussed here. The
action of macroscopic motions appears in the asymmetry of both E2
and E3. This gives explicit evidence of the influence of
macroscopic motions in observed PRD features, as advanced in
\cite{Carlin:2016aa}. 
Therefore it might be important to consider the action of dynamics
when studying the PRD polarization features of the second solar
spectrum.

Note that the PRD
minima of the eigenprofiles are broader and more separated than the minima produced by
dynamic Hanle. 
 Thus, the PCA analysis suggests that the dynamic Hanle
signature is in better agreement with the
width and location of the LP rings in observations. 
PCA does not explain the
large spatial scale of the LP rings.

Inspecting Fig. 4 of \cite{Trujillo-Bueno:2011aa}, we have detected Q/I and U/I rings in
quiet sun observations of another lower-chromosphere line: Ca {\sc ii} $8498$
{\AA}. In the same figure, the corresponding LP profiles of
the $\lambda8542$ and $\lambda8662$, which belong to the same triplet
but forms higher, exhibit slit patterns without rings. Instead, they are more like ``squared'' blocks along the
spatial direction of the slit, most of them of single sign. This
suggests that in observations of low temporal resolution the LP rings 
are favored by particular kinematic conditions of the
lower chromosphere, and that the
kinematic of upper layers somehow reduce their contrast and oval shapes. As PRD is
expected to be particularly negligible in $\lambda
8498$ \citep{Uitenbroek:1989aa}, this supports the
influence of Hanle and dynamics in the LP rings.

Our conclusion is that both
dynamics and PRD contribute to the near-core
sign-reversals, either because PRD itself is affected by dynamics at the very base of
the chromosphere or \textit{at least} because Hanle dynamics and PRD
signatures, though forming at different heights, overlap in wavelength 
in each timestep. 

 \begin{figure*}[h!]
\centering
\begin{center}$
\begin{array}{c}  
\includegraphics[width=0.84\textwidth]{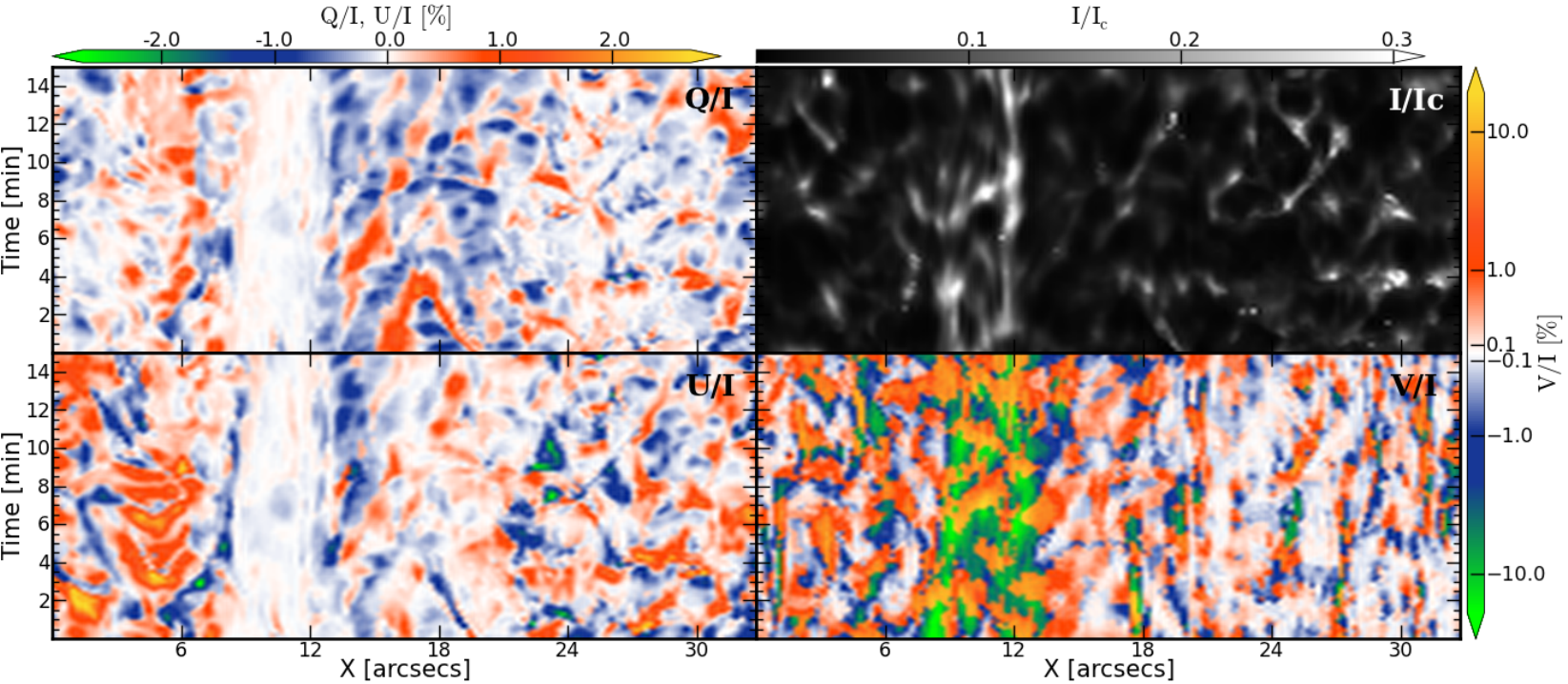} \cr  
\, \includegraphics[width=0.78\textwidth]{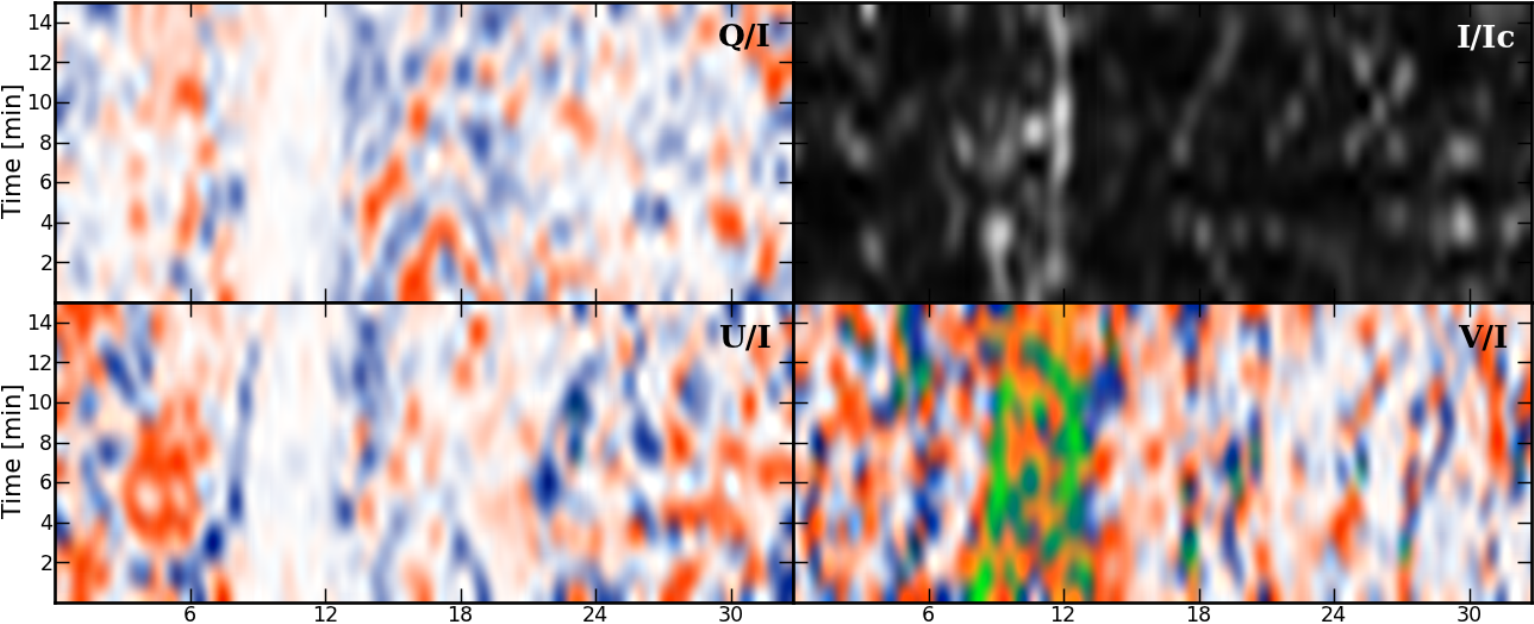}\quad\quad\quad\quad \cr  
\includegraphics[width=0.78\textwidth]{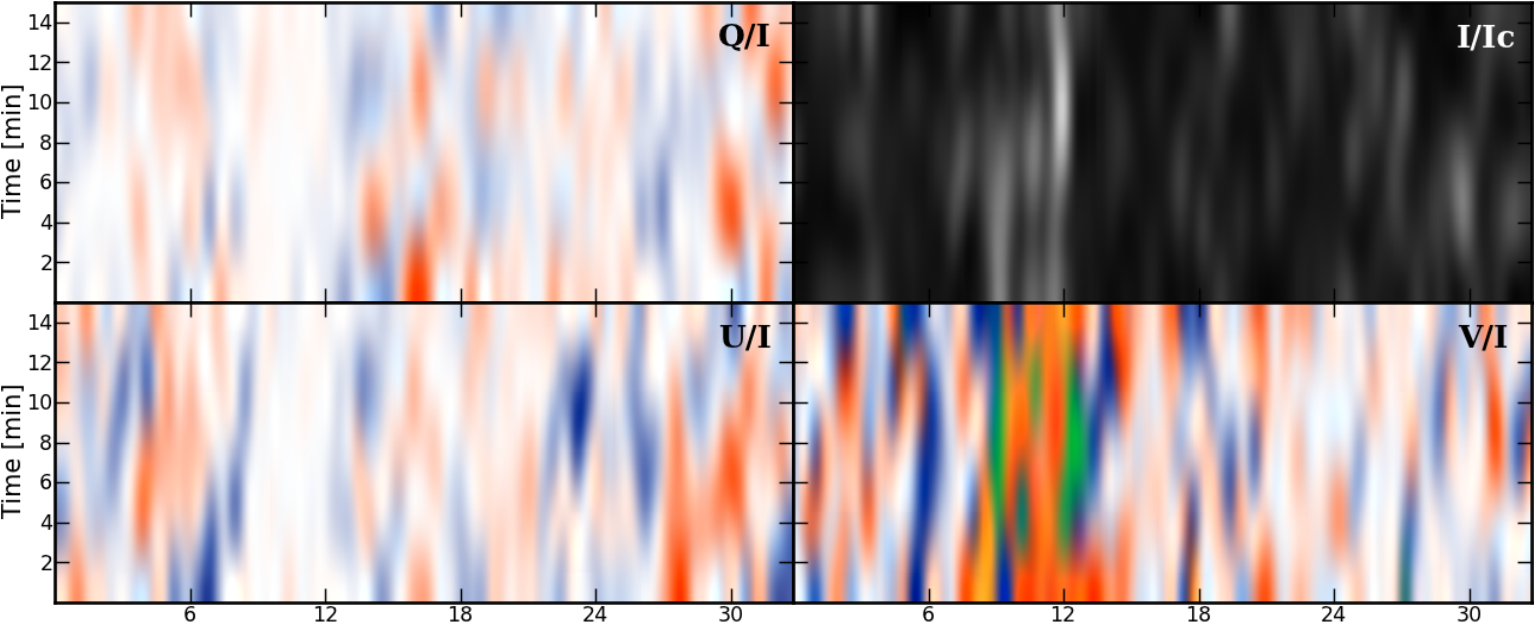}\quad\quad\quad\quad \cr  
\includegraphics[width=0.78\textwidth]{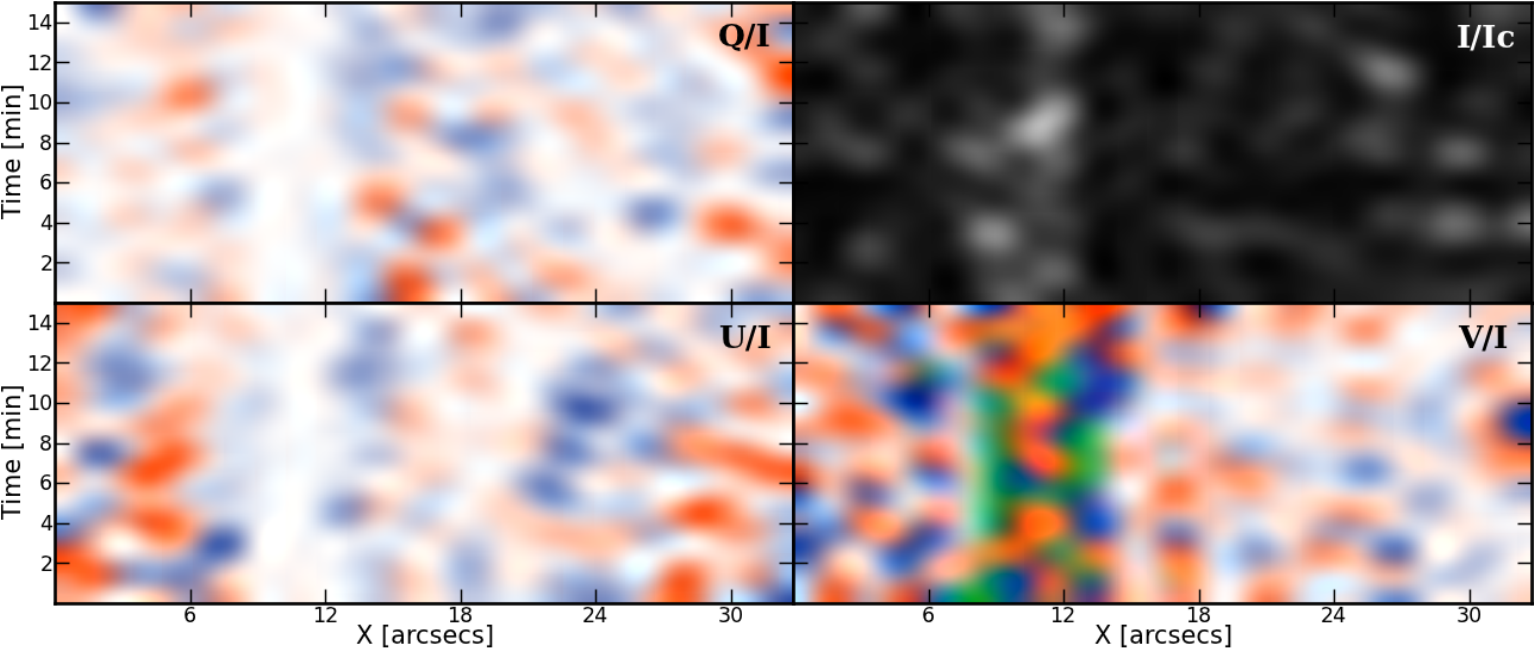} \quad\quad\quad\quad\\ 
\end{array}$
\end{center}
\vspace*{-5mm}
\caption{Time evolution of the slit Stokes vector for several resolutions. From top to bottom:
  $[\Delta \rm{x},\,\Delta \rm{ t}]$= $[0.2^{\prime\prime},10 s]$
 , $[0.4^{\prime\prime},60 s]$, $[0.4^{\prime\prime},180
  s]$, $[1.4^{\prime\prime},60 s]$. The polarization (intensity) values were chosen
in the maximum (minimum) of the spectral profile at each
location. Colorbars are common, the one for $V/I$ is logarithmic.}
\label{fig:xtmaps}
\end{figure*}

\subsection{Polarization time series}\label{sec:res2}
The study of the temporal evolution leads to
significant insights about the way the solar chromosphere generates
scattering polarization. First, because it 
exposes the effect of
dynamics, so giving the possibility of discriminating them to measure chromospheric
magnetic fields. Second, because it avoids the large degeneracy of
integrated signals, which can clarify the origin of the anomalies found in
the second solar spectrum.

Figure \ref{fig:xtmaps} shows spatio-temporal maps of fractional polarization 
at the wavelength of the absolute maximum of the
 profiles. This gives an estimation of the polarization structures
 that one may aim at observing with the S/N and 
spatio-temporal resolution of different ground solar
facilities (see
 caption of the figure\footnote{The degraded maps do not result from the direct
integration of the map in the top panel. In all the degraded
cases, the original signals of each Stokes component were
separately integrated in time,
space (in x and y directions of the slit) and wavelength but preserving the whole spectral profile in each
step. Then, the degraded maps were obtained by selecting the desired
wavelength.})
:
Irsol telescope (lower panels), Gregor (second group of panels from
below) and ATST-EST class telescope 
(third group of panels from below). The resolution of the top panels
is close to that of our calculations. 

These maps support the
existence of a sensitivity threshold, mentioned in the introduction, 
above which most of the (now hidden) scattering polarization signals
should appear `all at once' because the spatio-temporal scales of
chromospheric dynamics is resolved. 

\begin{figure}[t!]
\begin{center}$
\begin{array}{cc}
\includegraphics[scale=0.45]{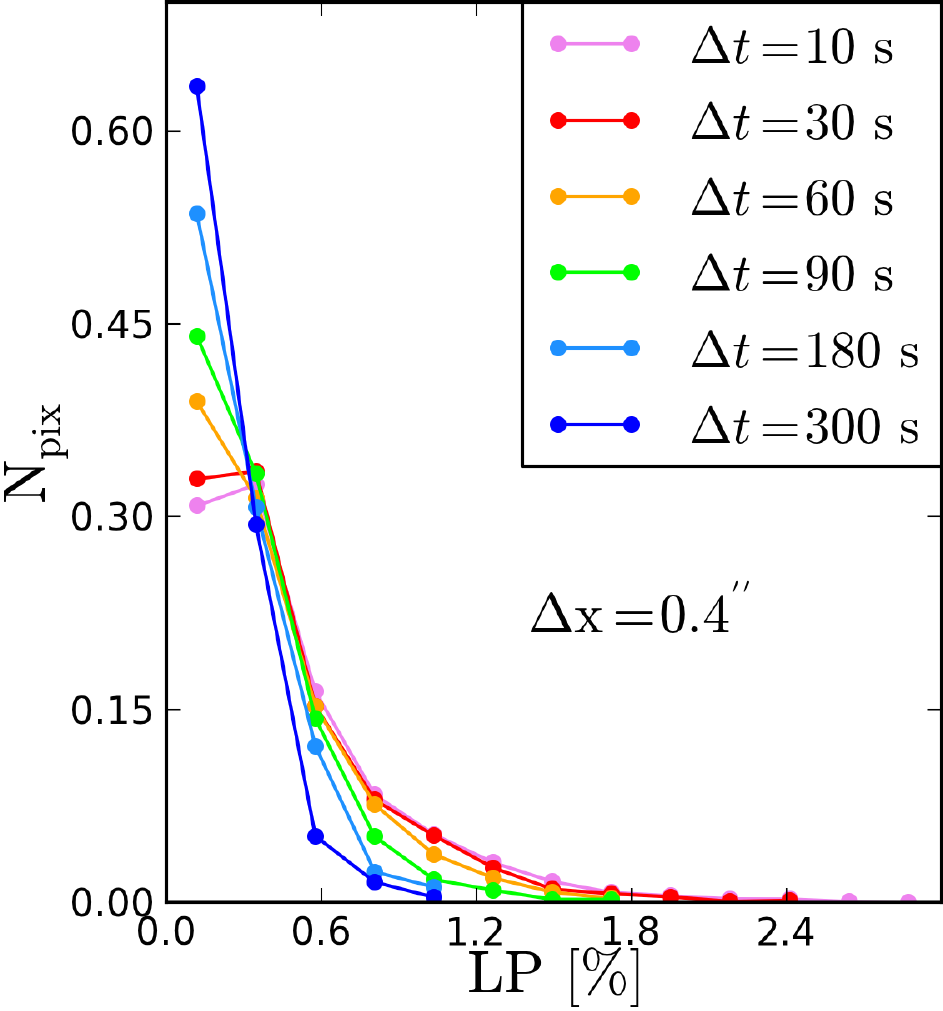} &    
\includegraphics[scale=0.45]{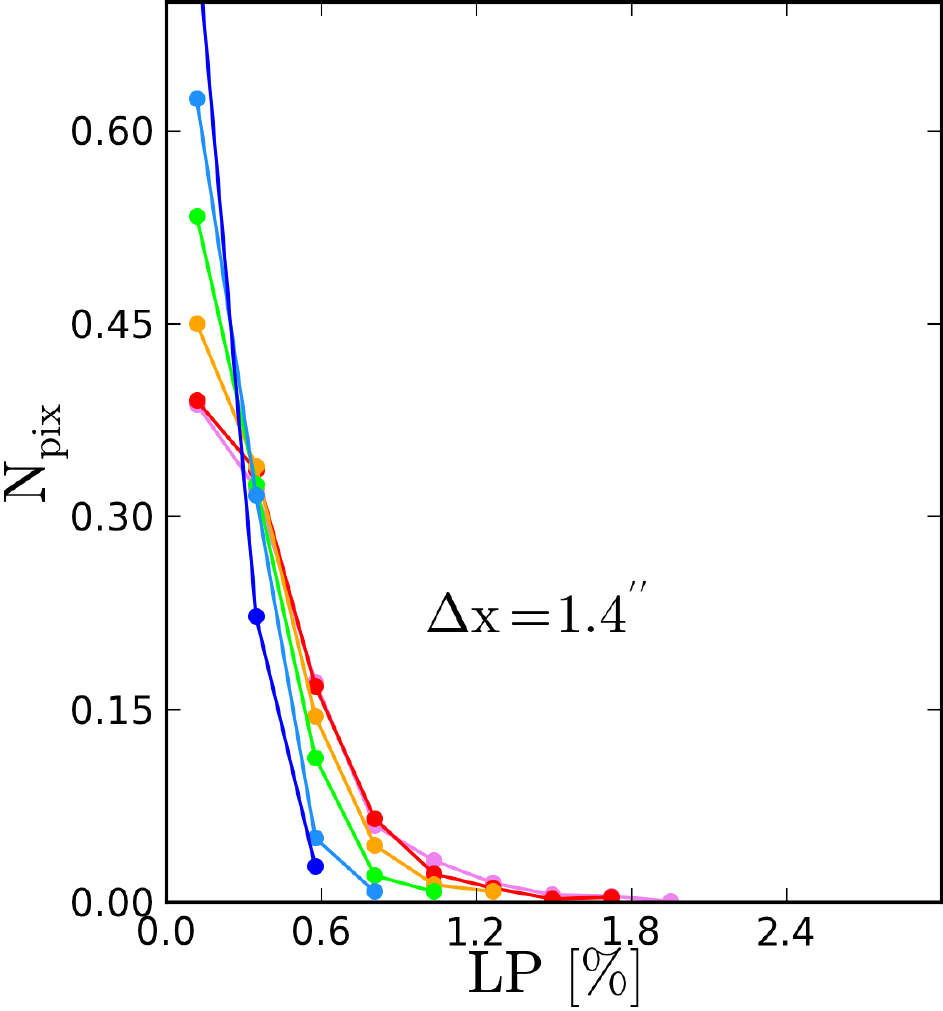} \\    
\includegraphics[scale=0.45]{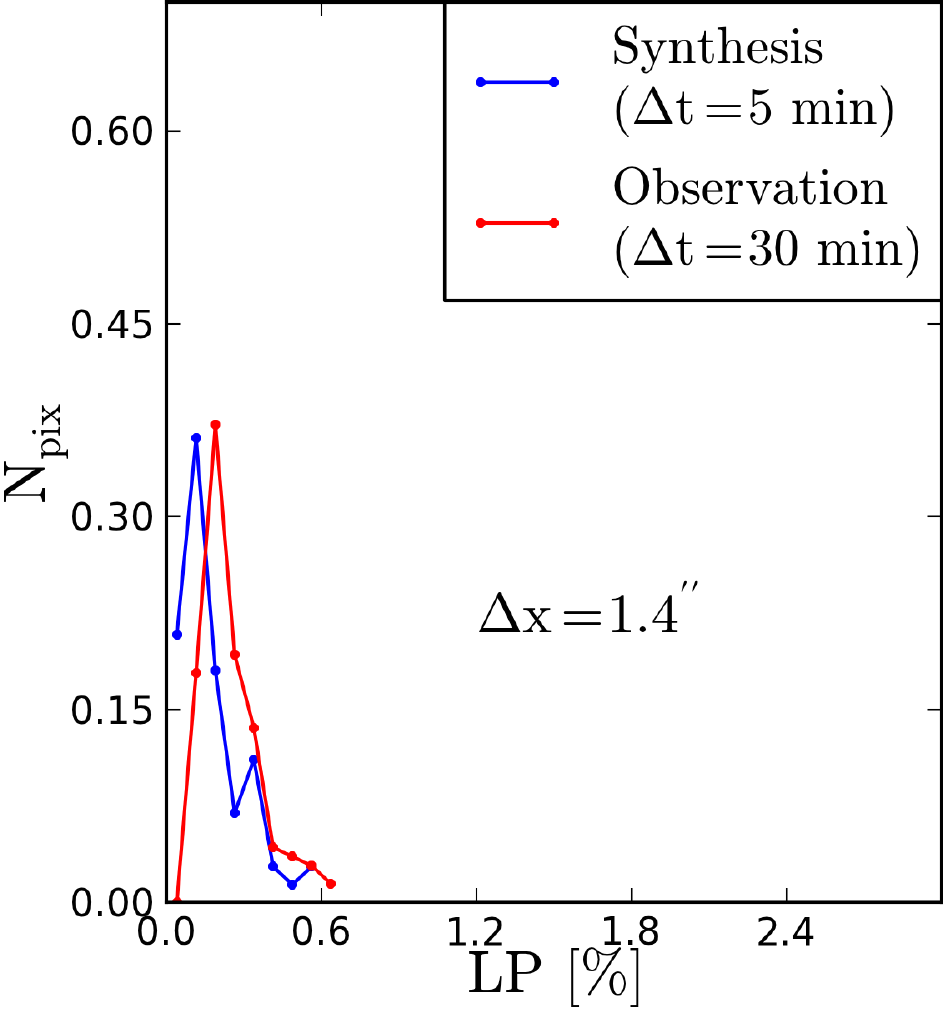} &
\includegraphics[scale=0.45]{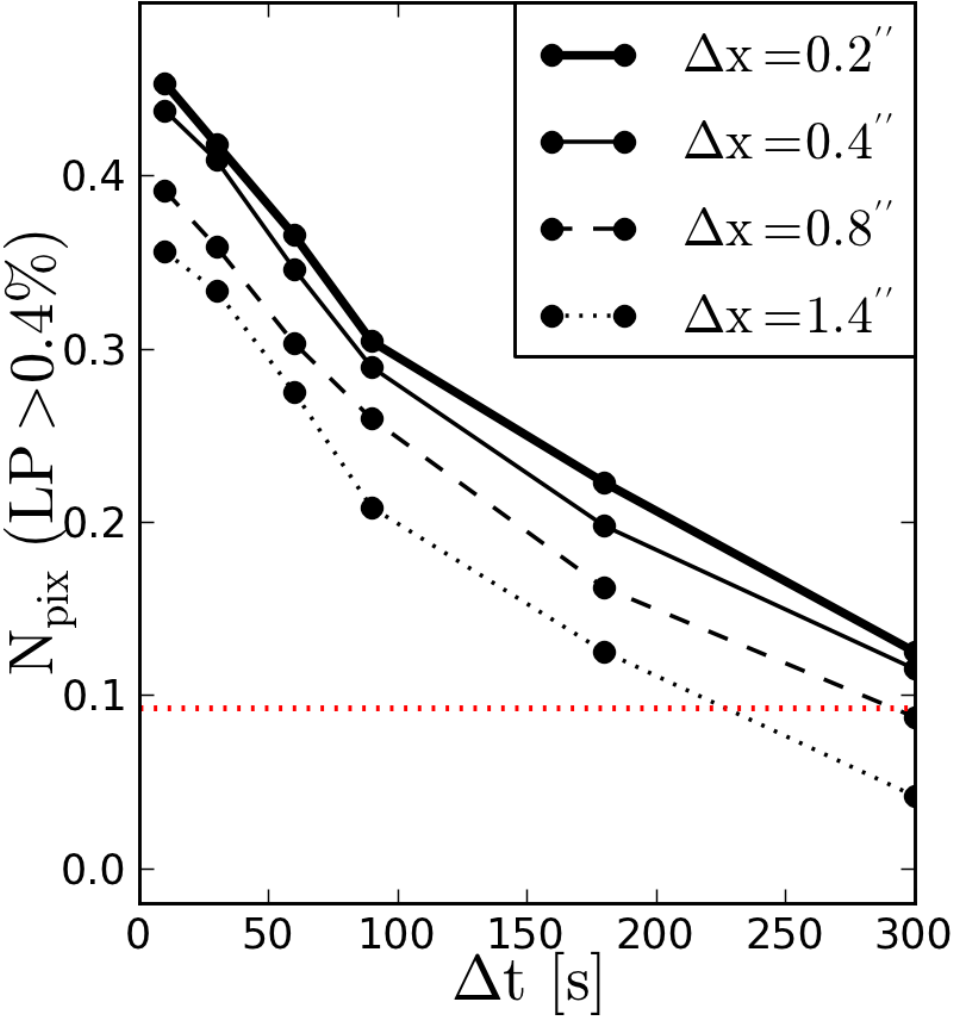} \\
\includegraphics[scale=0.45]{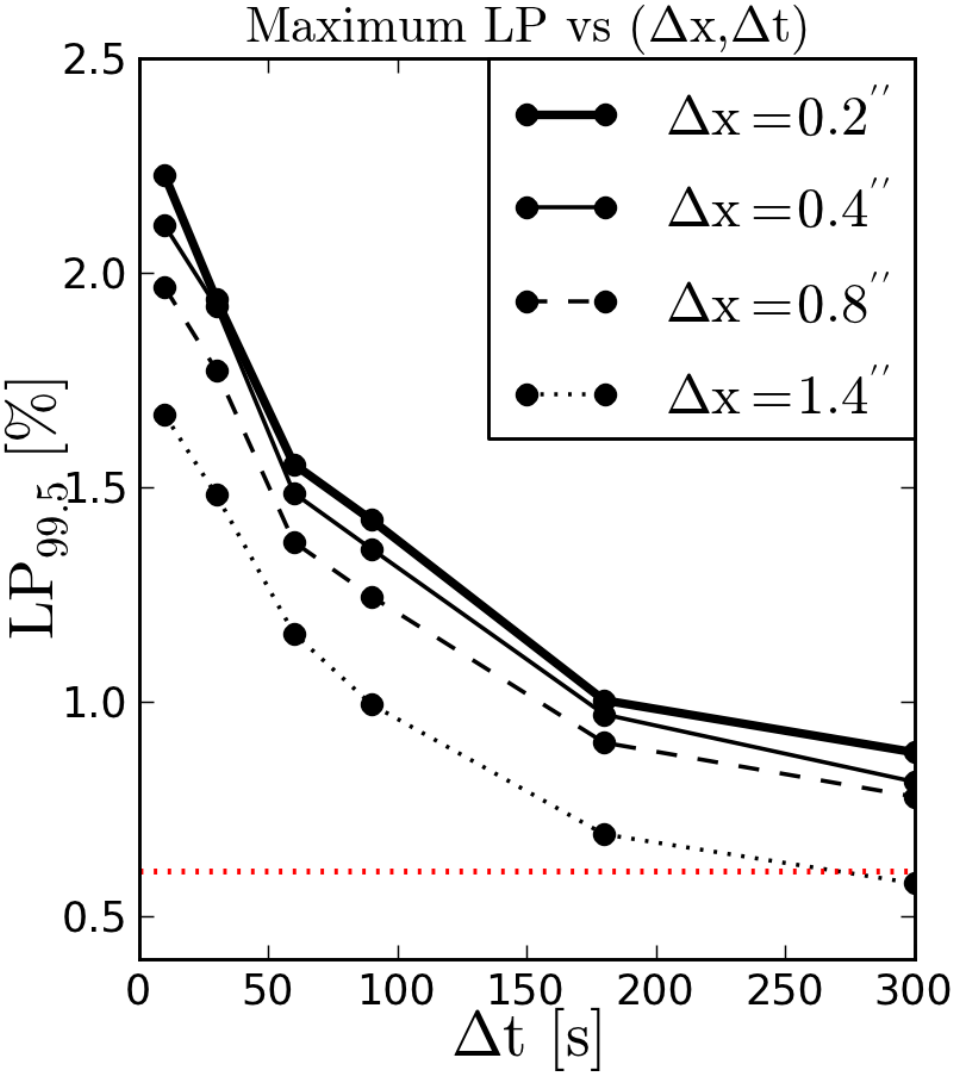}&
\includegraphics[scale=0.45]{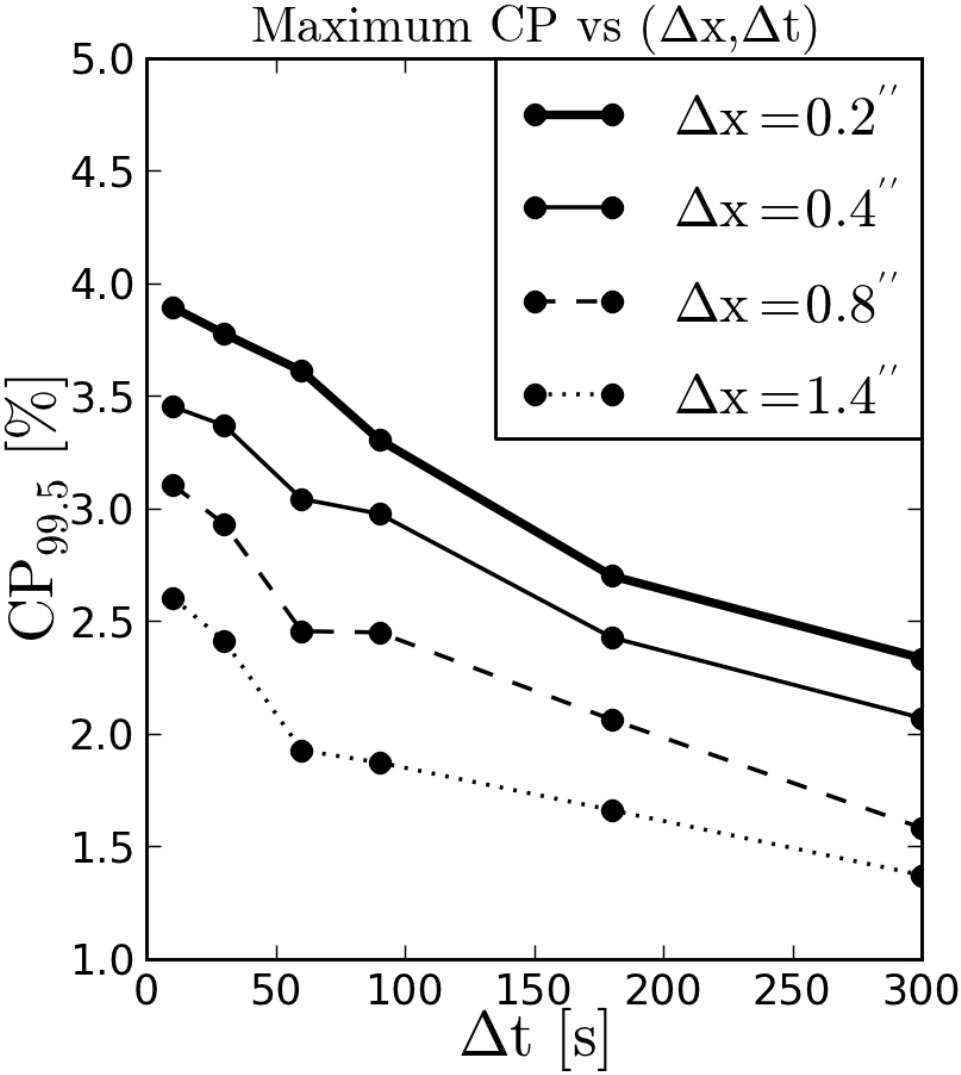} \\
\end{array}$

\end{center}
\caption{Upper panels: Envelopes of the normalized histograms for the
  total fractional LP in the given resolutions. The distributions cross
  in LP$\approx0.4\%$. Left middle panel: same dark blue curve as in
  upper right panel (but with higher bin resolution) compared to
  the observation.
 Right middle panel: relative (to one) number of pixels (area of
  the normalized histograms) with LP$>0.4\%$ versus the
  spatio-temporal resolution. Bottom panels: maximum fractional linear
(left) and circular (right) polarization quantified as the percentile
$99.5$. The horizontal red lines mark the corresponding value of the
LP in the observation.}
\label{fig:histo}
\end{figure}

Panels in Fig. \ref{fig:histo} quantify and characterize the variation of the polarization
amplitudes with the resolution. Note the different behavior for linear and circular
polarization. Q/I and U/I are more sensitive
to the temporal evolution of the atmosphere than V/I. Independently of the
spatial resolution, the first three
minutes of evolution produce the largest decay of LP, while such decay
is mild and almost linear in time for V/I. 

\section{Discussion}\label{sec:unexplained}
\subsection{Hanle diagnosis and polarization anomalies}
Our results show how the solar scattering
polarization can depend strongly on the evolution of the chromosphere. Note that despite the polarization amplitudes can decrease significantly with
integration time (Fig. \ref{fig:histo}), a given short interval with favorable
dynamics and magnetic fields can still restore the integrated LP to
 maximum values. This is so because the
instantaneous signals are intrinsically large for chromospheric kinematics and magnetic
fields and because a measurement is an integration, not an average. 
 
The spectral coherence of the polarization during the exposure
time matters. Assuming a typical period of $\approx3$ minutes for the
chromospheric evolution, we 
obtained reduced LP amplitudes of the order of $\approx \%1$
after that time (see panel of LP$_{99.5}$ in Fig. \ref{fig:histo}). This net $1$-period variation
depends on the balance between the kinematic amplification of
polarization due to the anisotropic Doppler brightenings \citep[][]{carlin12}, the lack of spectral
coherence along time (set by thermal broadening, Doppler shifts and
the phase between photosphere and chromosphere waves; see Letter),
 and the cancellations of LP due to transfer effects along the emergent rays. If the evolution
minimizes the fadings in a single-period and/or maximizes the amplifications
with sufficient regularity and spectral coherence over several periods, significant signals
can be measured after long exposures.  
The opposite situation can explain 
unexpectedly small signals in quiet sun locations where semiempirical
models predict LP well above the detection limit (see Sec. \ref{sec:hanle}).
Thus, other spectral lines and solar regions will have
their corresponding Fig. \ref{fig:histo}. 
We remark here the evolution of the magnetic field vector too. Its chromospheric inclination can
change during the emergence of shock-driven plasma bubbles. This
diminishes the LP intermittently, so reducing amplitudes
after $15$ minutes of integration (compare amplitudes in
$12<$x$<20$ Mm and x$> 20$ Mm in right-most panel of Fig. \ref{fig:figx1}).  
Through these mechanisms, dynamics increases the range of possible LP amplitudes.

This means that the systematic overlooking of kinematics and time
evolution has falsified the previous interpretations of
 chromospheric Hanle polarization. 
Curiously, such ``static'' approach appeals because both the lack of
resolution in the observations and the lack of kinematics in scattering
calculations compensate each other within the uncertainties that could
be explained with Hanle depolarization. Indeed, the maximum observed
polarization amplitudes of several spectral lines
agree with the maximum theoretical amplitudes given by
semiempirical, hence static, models of the chromosphere. 
This assertion is surprisingly precise for the Ca {\sc i} $4227$ {\AA} line because the maximum disk-center
total fractional LP amplitudes for any $B<150$ G are always between $0.4$-$0.5\%$(see Sec. \ref{sec:hanle}), very
close to the observed maximum of $0.6\%$. 
In cases without such an agreement, the differences have been 
typically explained assuming Hanle-depolarizing magnetic field inclinations, excluding the
suspicious and challenging anomalous excesses of line-core polarization
in certain chromospheric lines \citep{Landi-DeglInnocenti:1998aa,Stenflo:2000aa}. 
Our results point out that such difficult cases are precisely showing
the limitations of the static approach. 
The mere presence of shock waves is sufficient for
obtaining \textit{instantaneous} polarization enhancements of up to
one order of magnitude with respect to calculations in static or to temporally unresolved observations. Joining this with a coherent (constructive)
dynamics during long exposures, it is possible to explain the
excesses of polarization in the second solar spectrum. This
favorable situation should be more frequent at the limb, where the anomalous
signals appear. There, the height-dependent radiation
anisotropy tends to be mapped always in its positive part and the LOS Doppler shifts
produced by the emergent waves are minimized, so the
spectral LP enhancements are more easily coherent, hence reinforced.

The analysis and interpretation of the second solar spectrum requires
a new paradigm in which the origin of the scattering signals (the symphony) have to
be explained as the result of an atomic
system (the instrument) instantaneously reacting to the solar atmosphere (the playing
musician). A key remark is that \textit{the
 observed solar polarization is strongly determined by the temporal, geometrical and spectral
variation of the illumination that the scatterers receive}, which is ultimately
controlled by relative motions around them. Thus, polarization signals can
have both large variability and degeneracy, hence we need a way to
systematically expose, group or
constrain this broad range of possibilities. 
Can we do this with semiempirical models? Are MHD models enough? 

\subsection{About semiempirical models and heatings}\label{sec:models} 
The fact that semi-empirical models do not always explain
 the observed polarization has been noted by
 several works \citep[e.g., ][]{Smitha:2014aa, Supriya:2014aa}, though
 the reasons were unclear.
The dynamic effects affecting the scattering polarization cannot be
reproduced by semi-empirical models because they do not depend on time
and treat macroscopic
kinematics as ad-hoc microturbulent velocity. 
Dynamic effects cannot be modelled by macroturbulence neither, because
it just changes the shape of the emergent
signals but not the polarization properties of the media, similarly to
what happens with the velocity-free approximation
\citep{Carlin:2013aa}.

The effect of dynamics is
 more important as we increase resolution.
The microturbulent motions parameterized in semi-empirical
models act as effective Doppler shifts in the profiles when considering resolved observations or MHD simulations.
In other words, the dynamic ratio (Doppler
velocity in Doppler units),
\begin{equation*}
 \mathcal{\xi}({\rm s})=\frac{v_{\rm{res}}}{ v_{\rm{unres}} } = \frac{v_{\vec{\Omega}}}{\sqrt{v^2_{\rm{thermal}}+v^2_{\rm{micro}}}},
\end{equation*}
which controls the
modulation of anisotropy and LP,
increases with resolution because the microturbulent velocity
 ($v_{\rm{micro}}$) passes to be accounted in the
 numerator as resolved macroscopic velocity ($v_{\rm res}$=$v_{\vec{\Omega}}$) along the optical path ``s''
 of each ray $\vec{\Omega}$. 
 Note that unresolved velocities are both thermal and microturbulent, but only
 the latter gives a pool of motions that can act as effective Doppler shifts with increased resolution.

Larger $\mathcal{\xi}$ usually implies larger instantaneous LP
amplifications but, depending
on the evolution of the thermal broadening (heatings) and kinematics, it can also ease cancellations after
temporal integrations 
due to signs mixing and spectral incoherence. 
Thus, microturbulence not only compensates broadening (lack of
heating), it is also encoded in the LP amplitudes (Sec. \ref{sec:res1b}).
This happens through changes in the radiation field anisotropy
perceived by the scatterer but also in the amount of atomic population that is pumped to the
upper level. 
For instance, consider a scatterer in the coolest
solar layers. A broader absorption profile (larger microturbulence)
 allows it to capture more pumping light (even without motions) that otherwise would be screened by the
even broader profiles sandwiching the region of the temperature minimum. 
In summary, cooler plasma
means scattering polarization more sensitive to
kinematics and heatings. Therefore, the shape, width and
 amplitude of polarization profiles are a strong test/constrain for the models, even with time integration.
This and the current 
lack of understanding on the distribution of solar chromospheric
heatings mean that the choice of
microturbulence in calculations of polarization has larger relevance than usually believed.

 The fact that the chromospheric temperatures in the MHD models are cooler
 than in the sun \citep{Leenaarts:2009} motivated the MHD simulations with
 higher resolutions considered here.
But larger broadenings are still needed.
Extra broadening might come from
further-increased resolutions in the MHD calculation. However we think
that this will not improve 
the fit with the observations because
the LP profiles would be then narrower and with larger velocities, hence more
sensitive to the distribution of motions and prone to loose spectral
coherence. This would bring a more efficient
cancellation of LP (as in our results without microturbulence),
worsening the agreement with observations.
Thus, larger temperatures should come from purely thermal
sources ($v_{\rm{thermal}}$), not from unresolved motions.

Solving the chromospheric and coronal heating 
problems requires to understand how the heating sources are distributed in space and time in the solar plasma.
Remarkably this problem could be studied using spectral lines as the
$\lambda 4227$, whose
line core form in the lower chromosphere. 
There, with minimum
temperatures, small variations
in temperature become more notable than at the transition region or
corona,
at least in regard to scattering polarization. 

\subsection{Three-dimensional effects}
Though in some areas of the disk center the calculations in $1.5$D
and in $3$D seem to give similar results 
\citep{Stepan:2016aa}, the $1.5$D
approximation 
does not contain the effect of horizontal velocity gradients
nor of horizontal inhomogeneities.
The goal of the following discussion is to point out
 some subtleties related with the relative strength of each polarizing effect. 

The first one is that three-dimensional
effects in the solar polarization may
 be controlled by dynamics instead of by ``plasma
inhomogeneities'' (spatial lumps of
temperature and density).
The net pumping radiation at a given scattering point is affected by two contributions:
one given by the anisotropic radiation coming from
other points in an inhomogeneous \textit{but static} atmosphere;
and a latter one modulating the former when velocities
act in the radiative transfer connecting each point with the
scatterer. Namely, differential
velocities all over the formation region (seen by the scatterer) create 
opacity-changing Doppler shifts along the pumping rays
that change the radiation created by the inhomogeneities.
In particular, the distribution of horizontal kinematics is, as we
develop now, essential for understanding 3D effects because it
easily sculpts the effective horizontal radiation field.

The second idea follows. What happens in an atmosphere
without preferred horizontal directions? Here the light converging
horizontally in each
plasma element is affected, at each point of a long optical path, by \textit{randomly-organized horizontal velocity
gradients}. This can approximate a net cancellation of the positive and negative
azimuth-dependent radiation field components\footnote{This azimuthal
  isotropization of the field should be more effective the larger the
  horizontal extension of the 
  formation region and the less organized are the horizontal velocity
  gradients.}. In this way their contribution to the atomic polarization
can be largely exceeded by the combined
contributions of: 1) the ever-present and comparatively strong vertical gradients driven by
shock waves and gravity, which are geometrically ``organized'' ; and
2) the limb brightening/darkening. The $1.5$D approximation considers
both, hence if horizontal motions had such an isotropizing
  effect, the fractional LP in $1.5$D would approach the $3$D results
  in larger quiet sun areas. In this ``dynamic'' physical limit, disorganized horizontal velocity
 gradients minimize the azimuthal anisotropy of the radiation created by 
 inhomogeneities. Thus, though being an opposite situation to that of a
 $1.5$D atmosphere (limit without horizontal
 gradients/inhomogeneities), both situations let atomic polarization
 be driven by the vertical stratification of plasma properties. The solar atmosphere
  is somewhere between these two conceptual limits.

The last subtlety to mention has to do with the fact that three-dimensional
solar models are of relatively recent apparition. This implies that their chromospheric
 horizontal velocity fields lacks of observational feedback and should
 be expected to fail at reproducing the limb intensity of
 chromospheric lines. Contrarily to the ``observationally-tuned'' vertical
 velocities \citep{Carlsson:1997}, which are basically guided by
 shock waves and gravity, the less-investigated horizontal
 velocity (gradients) should furthermore depend on the distribution and scales of the magnetic fields
in the models. Adding this to the indirect sensitivity that the LP has on temperature
through kinematics (recall Sec. 4.2), we conclude that the
differences between $3$D and $1.5$D calculations might be quiet
influenced by ingredients of the models that require improvement.

We show the relevance of these ideas commenting 
on two of the few papers existing on this topic. The
key is that the influence of the
3D radiation field on the Hanle polarization has been
posed since the beggining in terms of inhomogeneities,
and not in terms of velocity gradients. 
The initial conclusions to this regard were obtained by Manso Sainz \&
Trujillo Bueno (2011) with an analytical/numerical
study approaching the problem in static regime. In more recent numerical
studies the action of 3D velocities in
the polarization is included, but the contributions of
 vertical and horizontal velocity gradients are not
 separated. As the effects of horizontal velocity gradients and
   inhomogeneities are in turn blended, it is also unknown which are
   their relative contributions to the differences between 1.5D and 3D
   calculations. Consider for instance Fig. 3 of \cite{Stepan:2016aa},
   which points out that the Hanle effect at disk center \textit{tends} to depolarize
when is accompanied by other symmetry breaking effect, as 
happens in the well-known case of the solar limb. The figure also confirms
that macroscopic motions are yet an efficient polarizing source for the spectral line
considered \citep[as known
from 1D simulations of the Ca
{\sc ii} IR triplet,][]{Carlin:2013aa}. 
However, it is unknown whether the polarization introduced by velocity
in that figure is due to 
the horizontal or to the vertical velocity gradients
 already considered in 1D calculations. This matters not only
for comparing with 1D calculations, but also because horizontal velocity
gradients can potentially 
compensate and exceed the polarization created by horizontal
inhomogeneities (as explained in previous paragraphs). Thus, the dominant
symmetry breaking and polarizing effect at disk center could not be the
inhomogeneities, as always afirmed, but the horizontal velocity gradients.
Hence, we think that is key to quantify the 
\textit{sensitivity} of such results to the \textit{model horizontal distribution of
velocity gradients}, which also can help to detach the conclusions
from the eventual lack of realism in them. Finally, it is
also to be noted that when those radiative transfer simulations do not add
microturbulent velocity, they use the
temperature distribution of MHD
 models that are known to represent a chromosphere cooler than the
solar one. Thus, as temperature influences significantly the effect of the velocities in
the LP (see Sec. 4.2 of the present paper), the relative strength of the polarizing
effects discussed here might change accordingly.


In summary,
horizontal velocity gradients
 compete with other physical mechanisms, such as inhomogeneities and
 Hanle effect, for polarizing and
depolarizing the light, hence we need to explicitly quantify whether they
are effectively relevant for inferring magnetic fields with the Hanle effect.
 We need to find theoretical and observational
 methods for discriminating the Hanle effect. Temporal evolution
 might help to this aim, which motivated the present paper. Other possibility is to explore the concept of Hanle PILs \citep[][]{Carlin:2015aa}: as in saturation the Hanle effect always nullify the polarization for particular magnetic field orientations,
 it creates an ever-present spatial fingerprint of the magnetic field topology in
 polarization maps.



\section{Conclusions}
Considering the radiation-MHD
simulation of \cite{Carlsson:2016aa}, we have simulated the temporal evolution of the spectral line
polarization of the Ca {\sc i} $4227$ {\AA} line in forward scattering, including the
Hanle and Zeeman effects. 

We find that the large forward-scattering amplitudes of the $\lambda 4227$
line are accentuated by its formation region, which is in the
temperature minimum and close to the
short-scale IN canopy
of horizontal magnetic fields. This maximizes the impact of macroscopic
motions in the LP and the Hanle effect respectively. 
 Without the amplification of polarization produced by
dynamics, the effect of temporal integration in current observations would make
impossible to detect the scattering polarization signals of this line
in the disk center. 

At least the strong
spectral lines forming in the lower chromosphere are expected to show
the largest sensitivity to 
kinematics and atmospheric heatings in linear polarization. Hence they
offer a possible
test bench for understanding the distribution of chromospheric
heatings through the scattering polarization. 
However, there where the evolution of dynamics has no sufficient spectral
and temporal coherence the cancellation of signals can make them
undetectable.
Thus, the evolution of dynamics 
might be key for explaining the polarization
profiles and perhaps the eventual absence of polarization in large areas.

Our calculations indicate that the measurement of
polarization time series with exposure times below a
minute should be of great help for Hanle diagnosis. In the Ca {\sc i} $4227$ {\AA} line, spatial
resolutions as large as $0.4^{\prime\prime}$ seem enough to detect the
Hanle structures of the chromosphere as soon as time resolution is achieved.

The discrimination of Hanle, Zeeman and PRD effects in the
generation of LP rings could not be completely clarified here but
the situation was exposed in better detail presenting several clues
that can guide deeper investigations. The near-core region of
the LP profiles is challenging because all possible effects overlap.

It seems necessary to revisit those studies of the second solar
spectrum where a formal comparison between theory and 
observations has been done without accounting for the
effect of kinematics and time evolution.

\acknowledgments 
Dedicated to Prof. Egidio Landi Degl'innocenti, who
 passed away during the publication of this paper.
We thank J. Leenaarts and M. Carlsson for providing the
atmospheric models and for valuable comments and discussions. We
thank J. Leenaarts for a careful reading of the manuscript that
helped to improve the text, and J. Stenflo and O. Steiner for their
feedback and discussions.   
This work was financed by the SERI project C12.0084 (COST action
MP1104) and by the Swiss National Science Foundation project $200021$\texttt{\_}$163405$.

\appendix
\begin{figure}[h!]
\centering
\includegraphics[scale=1]{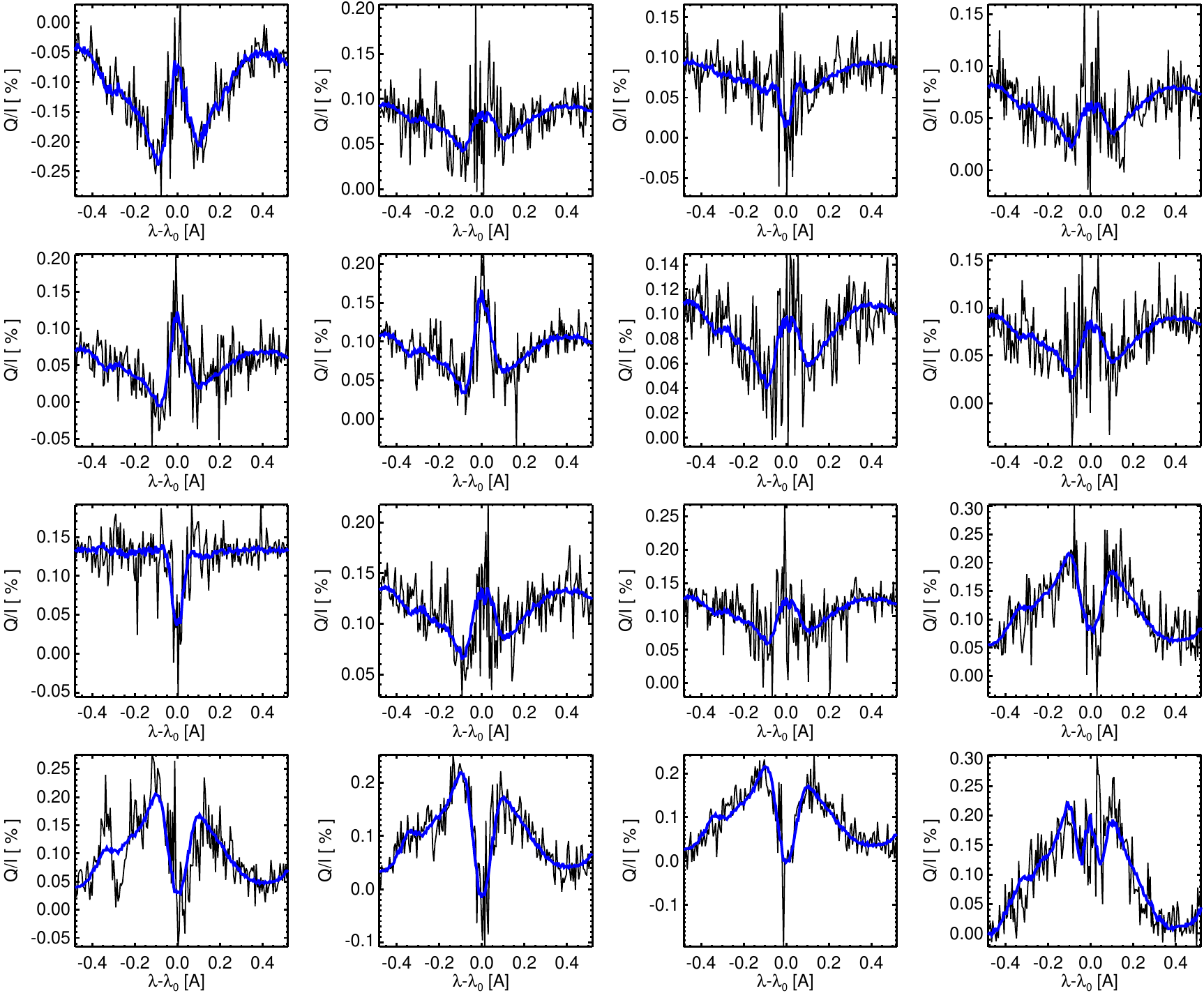}
\caption{Example of some random observed profiles of Ca {\sc i} $4227$
  {\AA} (black line) that have been reconstructed (blue line) using the first three PCA eigenvectors of our database.} 
\label{fig:pca2}
\end{figure}


\begin{thebibliography}{0}%
\makeatletter
\providecommand \@ifxundefined [1]{%
 \@ifx{#1\undefined}
}%
\providecommand \@ifnum [1]{%
 \ifnum #1\expandafter \@firstoftwo
 \else \expandafter \@secondoftwo
 \fi
}%
\providecommand \@ifx [1]{%
 \ifx #1\expandafter \@firstoftwo
 \else \expandafter \@secondoftwo
 \fi
}%
\providecommand \natexlab [1]{#1}%
\providecommand \enquote  [1]{``#1''}%
\providecommand \bibnamefont  [1]{#1}%
\providecommand \bibfnamefont [1]{#1}%
\providecommand \citenamefont [1]{#1}%
\providecommand \href@noop [0]{\@secondoftwo}%
\providecommand \href [0]{\begingroup \@sanitize@url \@href}%
\providecommand \@href[1]{\@@startlink{#1}\@@href}%
\providecommand \@@href[1]{\endgroup#1\@@endlink}%
\providecommand \@sanitize@url [0]{\catcode `\\12\catcode `\$12\catcode
  `\&12\catcode `\#12\catcode `\^12\catcode `\_12\catcode `\%12\relax}%
\providecommand \@@startlink[1]{}%
\providecommand \@@endlink[0]{}%
\providecommand \url  [0]{\begingroup\@sanitize@url \@url }%
\providecommand \@url [1]{\endgroup\@href {#1}{\urlprefix }}%
\providecommand \urlprefix  [0]{URL }%
\providecommand \Eprint [0]{\href }%
\providecommand \doibase [0]{http://dx.doi.org/}%
\providecommand \selectlanguage [0]{\@gobble}%
\providecommand \bibinfo  [0]{\@secondoftwo}%
\providecommand \bibfield  [0]{\@secondoftwo}%
\providecommand \translation [1]{[#1]}%
\providecommand \BibitemOpen [0]{}%
\providecommand \bibitemStop [0]{}%
\providecommand \bibitemNoStop [0]{.\EOS\space}%
\providecommand \EOS [0]{\spacefactor3000\relax}%
\providecommand \BibitemShut  [1]{\csname bibitem#1\endcsname}%
\let\auto@bib@innerbib\@empty
\end{thebibliography}%


\begin{thebibliography}{45}
\expandafter\ifx\csname natexlab\endcsname\relax\def\natexlab#1{#1}\fi

\bibitem[{{Anusha} {et~al.}(2011){Anusha}, {Nagendra}, {Bianda}, {Stenflo},
  {Holzreuter}, {Sampoorna}, {Frisch}, {Ramelli}, \& {Smitha}}]{Anusha:2011aa}
{Anusha}, L.~S., {Nagendra}, K.~N., {Bianda}, M., {Stenflo}, J.~O.,
  {Holzreuter}, R., {Sampoorna}, M., {Frisch}, H., {Ramelli}, R., \& {Smitha},
  H.~N. 2011, \apj, 737, 95

\bibitem[{{Anusha} {et~al.}(2010){Anusha}, {Nagendra}, {Stenflo}, {Bianda},
  {Sampoorna}, {Frisch}, {Holzreuter}, \& {Ramelli}}]{Anusha:2010aa}
{Anusha}, L.~S., {Nagendra}, K.~N., {Stenflo}, J.~O., {Bianda}, M.,
  {Sampoorna}, M., {Frisch}, H., {Holzreuter}, R., \& {Ramelli}, R. 2010, \apj,
  718, 988

\bibitem[{{Bianda} {et~al.}(2011){Bianda}, {Ramelli}, {Anusha}, {Stenflo},
  {Nagendra}, {Holzreuter}, {Sampoorna}, {Frisch}, \&
  {Smitha}}]{michele2011cai}
{Bianda}, M., {Ramelli}, R., {Anusha}, L.~S., {Stenflo}, J.~O., {Nagendra},
  K.~N., {Holzreuter}, R., {Sampoorna}, M., {Frisch}, H., \& {Smitha}, H.~N.
  2011, \aap, 530, L13

\bibitem[{{Bianda} {et~al.}(1998){Bianda}, {Solanki}, \&
  {Stenflo}}]{Bianda:1998aa}
{Bianda}, M., {Solanki}, S.~K., \& {Stenflo}, J.~O. 1998, \aap, 331, 760


\bibitem[{{Br{\"u}ckner}(1963)}]{bruckner1963}
{Br{\"u}ckner}, G. 1963, Z. Astrophys., 58

\bibitem[{{Carlin}(2015)}]{Carlin:2015ab}
{Carlin}, E.~S. 2015, in IAU Symposium, Vol. 305, Polarimetry, ed. K.~N.
  {Nagendra}, S.~{Bagnulo}, R.~{Centeno}, \& M.~{Jes{\'u}s Mart{\'{\i}}nez
  Gonz{\'a}lez}, 146--153

\bibitem[{{Carlin}(2016)}]{carlin:16b}
{Carlin}, E.~S. 2016, arXiv:1612.05091v2 [astro-ph.SR], SPW 8 proceedings

\bibitem[{Carlin \& {Asensio Ramos}(2015)}]{Carlin:2015aa}
Carlin, E.~S., \& {Asensio Ramos}, A. 2015, ApJ, 801, 16

\bibitem[{{Carlin} {et~al.}(2013){Carlin}, {Asensio Ramos}, \& {Trujillo
  Bueno}}]{Carlin:2013aa}
{Carlin}, E.~S., {Asensio Ramos}, A., \& {Trujillo Bueno}, J. 2013, \apj, 764,
  40

\bibitem[{{Carlin} \& {Bianda}(2016)}]{Carlin:2016aa}
{Carlin}, E.~S., \& {Bianda}, M. 2016, \apjl, 831, L5

\bibitem[{{Carlin} {et~al.}(2012){Carlin}, {Manso Sainz}, {Asensio Ramos}, \&
  {Trujillo Bueno}}]{carlin12}
{Carlin}, E.~S., {Manso Sainz}, R., {Asensio Ramos}, A., \& {Trujillo Bueno},
  J. 2012, ApJ, 751, 5

\bibitem[{{Carlsson} {et~al.}(2016){Carlsson}, {Hansteen}, {Gudiksen},
  {Leenaarts}, \& {De Pontieu}}]{Carlsson:2016aa}
{Carlsson}, M., {Hansteen}, V.~H., {Gudiksen}, B.~V., {Leenaarts}, J., \& {De
  Pontieu}, B. 2016, \aap, 585, A4

\bibitem[{{Carlsson} \& {Stein}(1997)}]{Carlsson:1997}
{Carlsson}, M., \& {Stein}, R.~F. 1997, ApJ, 481, 500


\bibitem[{{Collados} {et~al.}(2013){Collados}, {Bettonvil}, {Cavaller},
  {Ermolli}, {Gelly}, {P{\'e}rez}, {Soltau}, {Volkmer}, \& {EST
  Team}}]{Collados:2013aa}
{Collados}, M., {Bettonvil}, F., {Cavaller}, L., {Ermolli}, I., {Gelly}, B.,
  {P{\'e}rez}, H., A.~.-N., {Soltau}, D., {Volkmer}, R., \& {EST Team}. 2013,
  in Highlights of Spanish Astrophysics VII, 808--819

\bibitem[{{Dumont} {et~al.}(1973){Dumont}, {Pecker}, \&
  {Omont}}]{Dumont:1973aa}
{Dumont}, S., {Pecker}, J.-C., \& {Omont}, A. 1973, \solphys, 28, 271

\bibitem[{{Dumont} {et~al.}(1977){Dumont}, {Pecker}, {Omont}, \&
  {Rees}}]{Dumont:1977aa}
{Dumont}, S., {Pecker}, J.~C., {Omont}, A., \& {Rees}, D. 1977, \aap, 54, 675

\bibitem[{{Faurobert-Scholl}(1992)}]{Faurobert-Scholl:1992aa}
{Faurobert-Scholl}, M. 1992, \aap, 258, 521

\bibitem[{{Fontenla} {et~al.}(1993){Fontenla}, {Avrett}, \&
  {Loeser}}]{Fontenla:1993}
{Fontenla}, J.~M., {Avrett}, E.~H., \& {Loeser}, R. 1993, ApJ, 406, 319

\bibitem[{{Goodman}(1996)}]{Goodman:1996aa}
{Goodman}, M.~L. 1996, \apj, 463, 784

\bibitem[{{Gudiksen} {et~al.}(2011){Gudiksen}, {Carlsson}, {Hansteen}, {Hayek},
  {Leenaarts}, \& {Mart{\'{\i}}nez-Sykora}}]{Gudiksen:2011aa}
{Gudiksen}, B.~V., {Carlsson}, M., {Hansteen}, V.~H., {Hayek}, W., {Leenaarts},
  J., \& {Mart{\'{\i}}nez-Sykora}, J. 2011, \aap, 531, A154

\bibitem[{{Holzreuter} {et~al.}(2005){Holzreuter}, {Fluri}, \&
  {Stenflo}}]{Holzreuter:2005aa}
{Holzreuter}, R., {Fluri}, D.~M., \& {Stenflo}, J.~O. 2005, \aap, 434, 713

\bibitem[{{Lamb} \& {Ter Haar}(1971)}]{Lamb:1971aa}
{Lamb}, F.~K., \& {Ter Haar}, D. 1971, Physics Reports, 2, 253

\bibitem[{{Landi Degl'Innocenti}(1998)}]{Landi-DeglInnocenti:1998aa}
{Landi Degl'Innocenti}, E. 1998, \nat, 392, 256

\bibitem[{{Landi Degl'Innocenti} \& {Landolfi}(2004)}]{LL04}
{Landi Degl'Innocenti}, E., \& {Landolfi}, M. 2004, Polarization in Spectral
  Lines (Kluwer Academic Publishers)

\bibitem[{{Leenaarts} {et~al.}(2009){Leenaarts}, {Carlsson}, {Hansteen}, \&
  {Rouppe van der Voort}}]{Leenaarts:2009}
{Leenaarts}, J., {Carlsson}, M., {Hansteen}, V., \& {Rouppe van der Voort}, L.
  2009, \apjl, 694, L128

\bibitem[{{Leenaarts} {et~al.}(2012){Leenaarts}, {Pereira}, \&
  {Uitenbroek}}]{Leenaarts:2012aa}
{Leenaarts}, J., {Pereira}, T., \& {Uitenbroek}, H. 2012, \aap, 543, A109

\bibitem[{{Lites}(1974)}]{Lites:1974aa}
{Lites}, B.~W. 1974, \aap, 30, 297

\bibitem[{{Manso Sainz} \& {Trujillo Bueno}(2011)}]{manso11}
{Manso Sainz}, R., \& {Trujillo Bueno}, J. 2011, \apj, 743, 12

\bibitem[{{Mart{\'{\i}}nez Gonz{\'a}lez} {et~al.}(2008){Mart{\'{\i}}nez
  Gonz{\'a}lez}, {Asensio Ramos}, {Carroll}, {Kopf}, {Ram{\'{\i}}rez
  V{\'e}lez}, \& {Semel}}]{Martinez-Gonzalez:2008a}
{Mart{\'{\i}}nez Gonz{\'a}lez}, M.~J., {Asensio Ramos}, A., {Carroll}, T.~A.,
  {Kopf}, M., {Ram{\'{\i}}rez V{\'e}lez}, J.~C., \& {Semel}, M. 2008, \aap,
  486, 637

\bibitem[{{Pereira} \& {Uitenbroek}(2015)}]{Pereira:2015aa}
{Pereira}, T.~M.~D., \& {Uitenbroek}, H. 2015, \aap, 574, A3

\bibitem[{Ramelli {et~al.}(2010)Ramelli, Balemi, Bianda, Defilippis, Gamma,
  Hagenbuch, Rogantini, Steiner, \& Stenflo}]{Ramelli:2010aa}
Ramelli, R., Balemi, S., Bianda, M., Defilippis, I., Gamma, L., Hagenbuch, S.,
  Rogantini, M., Steiner, P., \& Stenflo, J.~O. 2010, SPIE, 77351Y

\bibitem[{{Rees} {et~al.}(2000){Rees}, {L{\'o}pez Ariste}, {Thatcher}, \&
  {Semel}}]{Rees:2000}
{Rees}, D.~E., {L{\'o}pez Ariste}, A., {Thatcher}, J., \& {Semel}, M. 2000,
  A\&A, 355, 759

\bibitem[{{Rimmele} {et~al.}(2013){Rimmele}, {Berger}, {McMullin}, {Keil},
  {Goode}, {Knoelker}, {Kuhn}, {Rosner}, {Casini}, {Lin}, {Woeger}, {von der
  Luehe}, {Tritschler}, \& {Atst Team}}]{Rimmele:2013aa}
{Rimmele}, T., {Berger}, T., {McMullin}, J., {Keil}, S., {Goode}, P.,
  {Knoelker}, M., {Kuhn}, J., {Rosner}, R., {Casini}, R., {Lin}, H., {Woeger},
  F., {von der Luehe}, O., {Tritschler}, A., \& {Atst Team}. 2013, in EGU
  General Assembly Conference Abstracts, Vol.~15, EGU General Assembly
  Conference Abstracts, 6305

\bibitem[{{Sampoorna} {et~al.}(2009){Sampoorna}, {Stenflo}, {Nagendra},
  {Bianda}, {Ramelli}, \& {Anusha}}]{Sampoorna:2009aa}
{Sampoorna}, M., {Stenflo}, J.~O., {Nagendra}, K.~N., {Bianda}, M., {Ramelli},
  R., \& {Anusha}, L.~S. 2009, \apj, 699, 1650

\bibitem[{{Skumanich} \& {L{\'o}pez Ariste}(2002)}]{Skumanich:2002}
{Skumanich}, A., \& {L{\'o}pez Ariste}, A. 2002, ApJ, 570, 379

\bibitem[{{Smitha} {et~al.}(2014){Smitha}, {Nagendra}, {Stenflo}, {Bianda}, \&
  {Ramelli}}]{Smitha:2014aa}
{Smitha}, H.~N., {Nagendra}, K.~N., {Stenflo}, J.~O., {Bianda}, M., \&
  {Ramelli}, R. 2014, \apj, 794, 30

\bibitem[{{Stenflo}(1974)}]{Stenflo:1974aa}
{Stenflo}, J.~O. 1974, \solphys, 37, 31

\bibitem[{{Stenflo} {et~al.}(2000){Stenflo}, {Keller}, \&
  {Gandorfer}}]{Stenflo:2000aa}
{Stenflo}, J.~O., {Keller}, C.~U., \& {Gandorfer}, A. 2000, \aap, 355, 789

\bibitem[{{Supriya} {et~al.}(2014){Supriya}, {Smitha}, {Nagendra}, {Stenflo},
  {Bianda}, {Ramelli}, {Ravindra}, \& {Anusha}}]{Supriya:2014aa}
{Supriya}, H.~D., {Smitha}, H.~N., {Nagendra}, K.~N., {Stenflo}, J.~O.,
  {Bianda}, M., {Ramelli}, R., {Ravindra}, B., \& {Anusha}, L.~S. 2014, \apj,
  793, 42

\bibitem[{{Tich{\'y}} {et~al.}(2015){Tich{\'y}}, {{\v S}t{\v e}p{\'a}n},
  {Trujillo Bueno}, \& {Kub{\'a}t}}]{Tichy:2015aa}
{Tich{\'y}}, A., {{\v S}t{\v e}p{\'a}n}, J., {Trujillo Bueno}, J., \&
  {Kub{\'a}t}, J. 2015, in IAU Symposium, Vol. 305, Polarimetry, ed. K.~N.
  {Nagendra}, S.~{Bagnulo}, R.~{Centeno}, \& M.~{Jes{\'u}s Mart{\'{\i}}nez
  Gonz{\'a}lez}, 401--406

\bibitem[{{Trujillo Bueno}(2001)}]{Trujillo-Bueno:2001aa}
{Trujillo Bueno}, J. 2001, in Astronomical Society of the Pacific Conference
  Series, Vol. 236, Advanced Solar Polarimetry -- Theory, Observation, and
  Instrumentation, ed. M.~{Sigwarth}, 161

\bibitem[{{Trujillo Bueno}(2011)}]{Trujillo-Bueno:2011aa}
{Trujillo Bueno}, J. 2011, in Astronomical Society of the Pacific Conference
  Series, Vol. 437, Solar Polarization 6, ed. J.~R. {Kuhn}, D.~M. {Harrington},
  H.~{Lin}, S.~V. {Berdyugina}, J.~{Trujillo-Bueno}, S.~L. {Keil}, \&
  T.~{Rimmele}, 83

\bibitem[{{Uitenbroek}(1989)}]{Uitenbroek:1989aa}
{Uitenbroek}, H. 1989, \aap, 213, 360

\bibitem[{{Uitenbroek}(2001)}]{Uitenbroek:2001}
---. 2001, ApJ, 557, 389

\bibitem[{{{\v S}t{\v e}p{\'a}n} \& {Trujillo Bueno}(2016)}]{Stepan:2016aa}
{{\v S}t{\v e}p{\'a}n}, J., \& {Trujillo Bueno}, J. 2016, \apjl, 826, L10

\bibitem[{{{\v S}t{\v e}p{\'a}n} {et~al.}(2015){{\v S}t{\v e}p{\'a}n},
  {Trujillo Bueno}, {Leenaarts}, \& {Carlsson}}]{Stepan:2015aa}
{{\v S}t{\v e}p{\'a}n}, J., {Trujillo Bueno}, J., {Leenaarts}, J., \&
  {Carlsson}, M. 2015, \apj, 803, 65

\end{thebibliography}

\end{document}